\documentclass[preprint,aps,12pt,showpacs,nofootinbib,tightenlines]{revtex4}
\usepackage{mathrsfs}
\usepackage{amsmath}
\usepackage{amssymb}
\usepackage{epsfig}
\usepackage{epstopdf}
\usepackage{graphicx}
\usepackage{subfigure}
\usepackage{booktabs}
\usepackage{float}
\usepackage{amsmath}
\usepackage{color}
\usepackage{multirow}

\def\be{\begin{eqnarray}}
\def\en{\end{eqnarray}}
\def\non{\nonumber\\}

\begin{document}
%%--------------------------------------------
\title{Study of $B_{(s)}$ meson decays to $D_{0}^{\ast}(2300) ,D_{s0}^{\ast}(2317) , D_{s1}(2460)$ and $D_{s1}(2536)$ within the covariant light-front approach}
\author{  You-Ya Yang, Zhi-Qing Zhang
\footnote{Electronic address: zhangzhiqing@haut.edu.cn (corresponding author)}, Hao Yang, Zhi-Jie Sun and Ming-Xuan Xie} %%
\affiliation{\it \small  Institute of Theoretical Physics, School of Sciences, Henan University
	of Technology, Zhengzhou, Henan 450001, China } %%
\date{\today}
\begin{abstract}
In this work, we investigate the form factors of the transitions $B_{(s)} \to D_{0}^{\ast}(2300),D_{s0}^{\ast}(2317),$ $ D_{s1}(2460) $ and $
D_{s1}(2536)$ in the covariant light-front quark model (CLFQM), where these final states are considered as P-wave excited charmed mesons. In order to obtain the form factors for the physical transition processes,
we extend these form factors from the space-like region to the time-like region. The $q^{2}$-dependence for each transition form factor is also plotted. Then, combined with those form factors, the branching ratios of the two-body nonleptonic decays $B_{(s)}\to D^*_{(s)0}(2300,2317)M, D_{s1}(2460,2536)M$ with $M$ being a light pseudoscalar (vector) meson or a charmed meson are calculated  by considering the QCD radiative corrections to the hadronic matrix elements within the QCD factorization approach. Most of our predictions are comparable to the results given by other theoretical approaches and the present available data.
\end{abstract}

 \vspace{1cm}

\maketitle

%=======================================================================
%                     Introduction
%=======================================================================

\section{Introduction}\label{intro}

In this work, we would like to study some of excited open-charm states, such as $D_{0}^{\ast}(2300),D_{s0}^{\ast}(2317), D_{s1}(2460)$ and $D_{s1}(2536)$ in $B_{(s)}$ meson decays. As we know, $D_{s0}^{\ast}(2317)$ and $D_{s1}(2460)$ were first discovered by BaBar \cite{babar} and CLEO \cite{cleo} in 2003, respectively. The scalar charm-strange meson $D_{s0}^{\ast}(2317)$ was observed in the invariant mass distribution of $D^+_s\pi^0$ and the axial charm-strange meson $D_{s1}(2460)$ was found in the invariant mass distribution of $D^{*+}_s\pi^0$. $D_{0}^{\ast}(2300)$ as the $SU(3)$ partner of $D_{0}^{\ast}(2317)$, called $D_{0}^{\ast}(2400)$ in the past, was discovered by Belle \cite{belle} in the three-body $B$ decay $B^+\to D^-\pi^+\pi^+$ in 2004. Another axial charm-strange meson with $J^P=1^+$ $D_{s1}(2536)$ was first observed in the $D^*_s\gamma$ invariant mass spectrum \cite{nun} in 1987.

Previous researches suggest that the masses of the resonants $D_{s0}^{\ast}(2317)$ and $ D_{s1}(2460)$ are just several tens MeV below the thresholds of $DK$ and $D^*K$, respectively. Furthermore, they are much lower than those given by the quark model \cite{quark}. Some abnormal  properties induce that these two states are difficult to be interpreted as conventional $c\bar s$ mesons. Therefore, many authors consider them as the $D^{(*)}K$ molecular states \cite{guo,cleven,close,guofk,lutz,lutz2,ylma}, the compact tetraquark states \cite{maiani,wangzg,hycheng2,yqchen,kim}, the chiral partners of the ground state $D^{(*)}_s$ mesons \cite{bardeen,nowak} and the states of $c\bar s$ mixed with four-quark states \cite{browder,vijande}. However, if considering the coupled channel effects and the fact that there are no additional states around the quark model predicted masses, these two states can be interpreted as P-wave charm-strange mesons \cite{bracco,lutz3,Hwang,xliu,xlchen,jbliu,wangzg2,fajfer,song,sfchen}. As to the $D_{0}^{\ast}(2300)$ state, 
besides the low mass puzzle, where the observed mass of $D_{0}^{\ast}(2300)$ \cite{belle, babar2} is bellow the predictions from the quark model about 100 MeV \cite{quark,godfrey2}, there exists the $SU(3)$ mass hierarchy puzzle, that is the masses of $D^*_0(2300)$ and $D^*_{s0}(2317)$ are almost equal to each other. These puzzles have triggered many studies on their inner structures: In Refs. \cite{bracco, nielsen}, the authors pointed out that the four-quark structure can explain the data measured by Belle \cite{belle} and BaBar \cite{babar2}, but not for that measured by FOCUS \cite{FOCUS:2003gru}. Another group authors solved these puzzles within the framework of unitarized chiral perturbation theory (UChPT) \cite{albala,mldu,guofk2} and considered that there exist two states in the $D^*_0(2300)$ energy region \cite{mldu2}, where the lighter one named as $D^*_0(2100)$ is the $SU(3)$ partner of the $D^*_{s0}(2317)$, the heavier one is a member of a different multiplet. Certainly, other authors also explained $D^*_0(2300)$ as a mixture of two and four-quark states \cite{vijande} or the bound state of $D\pi$ \cite{gamer}. The axial charm-strange meson $D_{s1}(2536)$ has been confirmed and studied in the $B_{(s)}$ meson decays by BaBar \cite{babar3}, Belle \cite{belle2} and LHCb \cite{lhcb}. Its measurements of mass and width are consistent with the theoretical expectations as a charm-strange meson with $J^P=1^+$. As we know that the QCD Lagrangian is invariant under the heavy flavor and spin rotation in the heavy quark limit. For the heavy mesons, the heavy quark spin $S_Q$ can decouple from the other degrees of freedom. Then $S_Q$ and the total angular momentum of the light quark $j$ become good quantum numbers. It is natural to label $D_{s1}(2460)$ and $D_{s1}(2536)$ by the quantum numbers $L^{j}_J$ with $j(L)$ being the total (orbital) angular momentum of the light quark, that is $P^{1/2}_1$ and $P^{3/2}_1$, denoted as $D^{1/2}_{s1}$ and $D^{3/2}_{s1}$, respectively. Since the heavy quark symmetry is not exact, the two $1^+$ states $D_{s1}(2460)$ and $D_{s1}(2536)$ can mix with each other through the following formula \cite{belle3}
\begin{eqnarray}
	|D_{s 1}(2460)\rangle &=&|D_{s 1}^{1 / 2}\rangle  \sin \theta_{s}+|D_{s 1}^{3 / 2}\rangle  \cos \theta_{s}, \nonumber\\
	|D_{s 1}(2536)\rangle &=&-|D_{s 1}^{3 / 2}\rangle  \sin \theta_{s}+|D_{s 1}^{1 / 2}\rangle  \cos \theta_{s}.
	\label{mixing1}
\end{eqnarray} 
While the states $D_{s 1}^{1/2}$ and $D_{s 1}^{3/2}$ are expected to be a mixture of states $^1D_{s1}$ and $^3D_{s1}$ with $J^{PC}=1^{++}$ and $1^{+-}$, respectively,
\begin{eqnarray}
	|D^{3/2}_{s1}\rangle &=& \sqrt{\frac{2}{3}} |^1D_{s1}\rangle +\sqrt{\frac{1}{3}}
	|^3D_{s1}\rangle,\nonumber\\
	|D^{1/2}_{s1}\rangle &=& -\sqrt{\frac{1}{3}} |^1D_{s1}\rangle +\sqrt{\frac{2}{3}}
	|^3D_{s1}\rangle.\label{mixing2}
\end{eqnarray}
Combining Eq. (\ref{mixing1}) and Eq. (\ref{mixing2}), one can find that
\be
|D_{s 1}(2460)\rangle &=& |^1D_{s1}\rangle \cos \theta+|^3D_{s1}\rangle\sin \theta , \nonumber\\
|D_{s 1}(2536)\rangle &=& -|^1D_{s1}\rangle \sin \theta+|^3D_{s1}\rangle\cos \theta.
\label{mixing3}
\en
where $\theta=\theta_s+35.3^\circ$ \cite{zhw}.

Using the manifestly covariant of the Bethe-Salpeter (BS) approach \cite{Salpeter:1951sz,Salpeter:1952ib}, Jaus, Choi, Ji and Cheng \emph{et al}. \cite{Jaus:1999zv,Choi:1998nf,Cheng:1997au} put forward the CLFQM around 2000. This approach provides a systematic way to explore the zero-mode effects, which are just cancelled by involving the spurious contributions being proportional to the lightlike four-vector $\omega=(0,2,0_{\bot})$, at the same time the covariance of the matrix elements being restored \cite{Jaus:1999zv}. Up to now the CLFQM has been used extensively to study the weak and radiative decays, as well as the features of some exotic hadrons \cite{Cheng:2003sm,w.wang,x.wang,w.wang2,ke,Li:2010bb,Sun1,Sun2,Sun3}. In this work, we will employ the CLFQM to evaluate the $B_{(s)}\to D_{0}^{\ast}(2300), D_{s0}^{\ast}(2317), D_{s1}(2460), D_{s1}(2536)$ transition form factors, then calculate the branching ratios of the relevant decays. In our calculations these hadrons are regarded as ordinary meson states. Compared with the future experimental measurements, our predictions are helpful to clarify the inner structures of these four hadrons.

The arrangement of this paper is as follows: In Section II, an introduction to the CLFQM and the expressions for the form factors of the transitions $B_{(s)}\to D_{0}^{\ast}(2300), D_{s0}^{\ast}(2317), D_{s1}(2460), D_{s1}(2536)$  are presented. Then the branching ratios of the $B_{(s)}$ meson decays with one of these considered P-wave excited charm sates involved are calculated under the QCD factorization approach, where the vertex corrections and the hard spectator-scattering corrections are considered. In Section III, the numerical results of the transition form factors and their $q^2$-dependence are presented. Then, combined with the transition form factors, the branching ratios of the decays $B_{(s)}\to D^*_{(s)0}(2300,2317)M, D_{s1}(2460,2536)M$ with $M$ being a light pseudoscalar (vector) meson or a charmed meson are calculated. In addition, detailed numerical analysis and discussion, including comparisons with the data and other model calculations, are carried out. The conclusions are presented in the final part.

\section{Formalism}\label{form}
\subsection{The Covariant Light-Front Quark Model}
Under the covariant light-front quark model, the light-front coordinates of a momentum $p$ are defined as $p=(p^-,p^+,p_\perp)$ with
$p^\pm=p^0\pm p_z$ and $p^2=p^+p^--p^2_\perp$. If the momenta of the quark and antiquark
with mass $m_{1}^{\prime(\prime\prime)}$ and $m_2$ in the incoming (outgoing) meson are denoted as $p_{1}^{\prime(\prime\prime)}$
and $p_{2}$, respectively, the momentum of the incoming (outgoing) meson with mass $M^\prime(M^{\prime\prime})$ can be written as $P^\prime=p_1^\prime+p_2 (P^{\prime\prime}=p_1^{\prime\prime}+p_2)$.
Here, we use the same notation
as those in Refs. \cite{Jaus:1999zv,Cheng:2003sm} and $M^\prime$ refers to $m_{B}$ for $B$ meson decays.
These momenta can be related each other through the internal variables $(x_{i},p{'}_{\perp})$
\be
p_{1,2}^{\prime+}=x_{1,2} P^{\prime+}, \quad p_{1,2 \perp}^{\prime}=x_{1,2} P_{\perp}^{\prime} \pm p_{\perp}^{\prime},
\en
with $x_{1}+x_{2}=1$. Using these internal variables,
we can define some quantities for the incoming meson which will be used in the following calculations:
\be
M_{0}^{\prime 2} &=&\left(e_{1}^{\prime}+e_{2}\right)^{2}=\frac{p_{\perp}^{\prime 2}+m_{1}^{\prime 2}}{x_{1}}
+\frac{p_{\perp}^{2}+m_{2}^{2}}{x_{2}}, \quad \widetilde{M}_{0}^{\prime}=\sqrt{M_{0}^{\prime 2}-\left(m_{1}^{\prime}-m_{2}\right)^{2}}, \non
e_{i}^{(\prime)} &=&\sqrt{m_{i}^{(\prime) 2}+p_{\perp}^{\prime 2}+p_{z}^{\prime 2}}, \quad \quad p_{z}^{\prime}
=\frac{x_{2} M_{0}^{\prime}}{2}-\frac{m_{2}^{2}+p_{\perp}^{\prime 2}}{2 x_{2} M_{0}^{\prime}},\en
where the kinetic invariant mass of the incoming meson $M'_0$ can be expressed as the energies of the quark and the antiquark
$e^{(\prime)}_i$. It is similar to the case of the outgoing meson.

\begin{figure}[htbp]
\centering \subfigure{
\begin{minipage}{5cm}
\centering
\includegraphics[width=5cm]{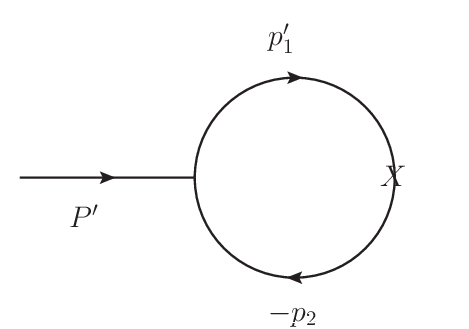}
\end{minipage}}
\subfigure{
\begin{minipage}{6cm}
\centering
\includegraphics[width=6cm]{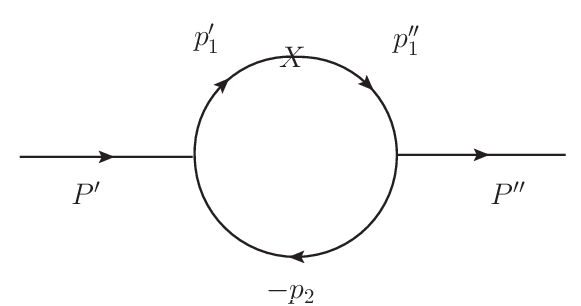}
\end{minipage}}
\caption{Feynman diagrams for $B$ decay (left) and transition
(right) amplitudes, where $P^{\prime(\prime\prime)}$ is the
incoming (outgoing) meson momentum, $p^{\prime(\prime\prime)}_1$
is the quark momentum, $p_2$ is the antiquark momentum and X
denotes the vector or axial vector transition vertex.}
\label{feyn}
\end{figure}

The form factors of the transitions
$B\to D^*_0$ \footnote{It is similar for the transition $B\to D^*_{s0}$. From now on, we will use $D^*_0$ and $D^*_{s0}$ to represent $D^*_0(2300)$ and $D^*_{s0}(2317)$, respectively, for simplicity.} and $B\to \ ^iD_{s1} (i=1, 3)$ \footnote{We will use $D_{s1}$ and $D_{s1}^{\prime}$ to represent
$D_{s1}(2460)$ and $D_{s1}(2536)$, respectively. The form factors of the transitions $B\to D_{s1}$ and $B\to D_{s1}^{\prime}$ can be obtained from those of the transitions $B\to \ ^1D_{s1}$ and $B\to \ ^3D_{s1}$ through Eq. (\ref{mixing3}). } induced by the vector and aixal-vector currents
are defined as
\be
\left\langle D^*_0 \left(P^{\prime
\prime}\right)\left|A_{\mu}\right|
B\left(P^{\prime}\right)\right\rangle&=&i\left[u_+(q^2)P_\mu+u_-(q^2)q_\mu\right],\\
\left\langle \; ^iD_{s1} \left(P^{\prime
\prime},\varepsilon\right)\left|A_{\mu}\right|
B\left(P^{\prime}\right)\right\rangle&=-&q(q^2)\epsilon_{\mu\nu\alpha\beta}\varepsilon^{*\nu}P^\alpha q^\beta,\\
\left\langle \; ^iD_{s1}\left(P^{\prime
\prime},\varepsilon\right)\left|V_{\mu}\right|
B\left(P^{\prime}\right)\right\rangle&=&i\left\{l(q^2)\varepsilon_\mu^*+\varepsilon^*\cdot P\left[P_\mu c_+(q^2)+q_\mu c_-(q^2)\right]\right\}.
\en
In calculations, the Bauer-Stech-Wirbel (BSW) \cite{bsw} transition form factors are more frequently used and defined by
\begin{footnotesize}
\begin{eqnarray}
\left\langle D^*_0 \left(P^{\prime
	\prime}\right)\left|A_{\mu}\right|
B\left(P^{\prime}\right)\right\rangle&=&\left(P_{\mu}-\frac{m_{B}^{2}-m_{D^*_0}^{2}}{q^{2}}
q_{\mu}\right) F_{1}^{B
	D^*_0}\left(q^{2}\right)+\frac{m_{B}^{2}-m_{D^*_0}^{2}}{q^{2}} q_{\mu}
F_{0}^{B D^*_0}\left(q^{2}\right),\;\;\;\;\;\;\;\;\\
\left\langle ^iD_{s1} \left(P^{\prime \prime}, \varepsilon^{\mu *}\right)\left|V_{\mu}\right| B\left(P^{\prime}\right)\right\rangle
&=&-i\left\{\left(m_{B}-m_{^iD_{s1}}\right) \varepsilon_{\mu}^{ *} V_{1}^{B \ ^iD_{s1}}\left(q^{2}\right)
-\frac{\varepsilon^{ *} \cdot P}{m_{B}-m_{^iD_{s1}}} P_{\mu} V_{2}^{B \ ^iD_{s1}}\left(q^{2}\right)\right.\non &&
\left.-2 m_{^iD_{s1}} \frac{\varepsilon^{ *} \cdot
P}{q^{2}} q_{\mu}\left[V_{3}^{B \
^iD_{s1}}\left(q^{2}\right)-V_{0}^{B \
^iD_{s1}}\left(q^{2}\right)\right]\right\},\label{bctoa1}\\
\left\langle ^iD_{s1}\left(P^{\prime \prime}, \varepsilon^{\mu*}\right)\left|A_{\mu}\right| B\left(P^{\prime}\right)\right\rangle
&=&-\frac{1}{m_{B}-m_{^iD_{s1}}} \epsilon_{\mu \nu \alpha \beta} \varepsilon^{ * \nu} P^{\alpha} q^{\beta} A^{B\ ^iD_{s1}}\left(q^{2}\right),
\end{eqnarray}
\end{footnotesize}
where $P=P'+P'', q=P'-P''$, and the convention $\epsilon_{0123}=1$ is adopted.

To smear the singularity at $q^2=0$ in Eq. (\ref{bctoa1}), the relations $V^{B \ ^iD_{s1}}_3(0)=V^{B\ ^iD_{s1}}_0(0)$ are required, and
\be
V^{B\ ^iD_{s1}}_3(q^2)=\frac{m_{B}-m_{^iD_{s1}}}{2m_{^iD_{s1}}}V^{B\ ^iD_{s1}}_1(q^2)-\frac{m_{B}+m_{^iD_{s1}}}{2m_{^iD_{s1}}}V^{B\ {^iD_{s1}}}_2(q^2).
\en
These two kinds of form factors are related to each other via
\be
F^{BD^*_0}_1(q^2)&=&-u_+(q^2),F^{BD^*_0}_0(q^2)=-u_+(q^2)-\frac{q^2}{q\cdot P}u_-(q^2),\label{relations}\\
A^{B\ ^iD_{s1}}(q^2)&=&-(m_{B}-m_{^iD_{s1}})q(q^2), V^{B\ ^iD_{s1}}_1(q^2)=-\frac{l(q^2)}{m_{B}-m_{^iD_{s1}}},\label{relationa1}\\
V^{B\ ^iD_{s1}}_2(q^2)&=&(m_{B}-m_{^iD_{s1}})c_+(q^2),V^{B\ ^iD_{s1}}_3(q^2)-V^{B\ ^iD_{s1}}_0(q^2)=\frac{q^2}{2m_{^iD_{s1}}}c_-(q^2).\label{relationa2}
\en
The light-front wave functions (LFWFs) are needed in the form factor calculations. Although the LFWFs can be derived from solving the relativistic Schr$\ddot{o}$dinger equation theoretically, it is difficult to obtained their exact solutions in many cases. Consequently, we will use the phenomenological Gaussian-type wave functions in this work,
\be
\varphi^{\prime} &=&\varphi^{\prime}\left(x_{2}, p_{\perp}^{\prime}\right)=4\left(\frac{\pi}{\beta^{\prime 2}}\right)^{\frac{3}{4}}
\sqrt{\frac{d p_{z}^{\prime}}{d x_{2}}} \exp \left(-\frac{p_{z}^{\prime 2}+p_{\perp}^{\prime 2}}{2 \beta^{\prime 2}}\right),\non
\varphi_{p}^{\prime} &=&\varphi_{p}^{\prime}\left(x_{2}, p_{\perp}^{\prime}\right)=\sqrt{\frac{2}{\beta^{\prime 2}}} \varphi^{\prime},
\quad \frac{d p_{z}^{\prime}}{d x_{2}}=\frac{e_{1}^{\prime} e_{2}}{x_{1} x_{2} M_{0}^{\prime}},\label{betap}
\en
where the parameter $\beta'$ describes the momentum distribution and is approximately of order $\Lambda_{QCD}$. It can be usually determined by the decay constants through the following analytic expressions \cite{Jaus:1999zv,Cheng:2003sm},
\be
f_{D^*_0}&=&\frac{N_{c}}{16 \pi^{3}} \int d x_{2} d^{2} p_{\perp}^{\prime} \frac{h_{D^*_0}^{\prime}}{x_{1} x_{2}\left(M^{\prime 2}-M_{0}^{\prime 2}\right)}
4\left(m_{1}^{\prime} x_{2}-m_{2} x_{1}\right),\\
f_{\;^3D_{s1}}&=&-\frac{N_{c}}{4 \pi^{3} M^{\prime}}  \int d x_{2} d^{2} p_{\perp}^{\prime} \frac{h_{^3D_{s1}}^{\prime}}{x_{1} x_{2}\left(M^{\prime 2}-M_{0}^{\prime 2}\right)}\non &&
\times\left[x_{1} M_{0}^{\prime 2}-m_{1}^{\prime}\left(m_{1}^{\prime}+m_{2}\right)-p_{\perp}^{\prime 2}-\frac{m_{1}^{\prime}-m_{2}}{w_{\;^3D_{s1}}^{\prime}} p_{\perp}^{\prime 2}\right],\\
f_{\;^1D_{s1}}&=&\frac{N_{c}}{4 \pi^{3} M^{\prime}} \int d x_{2} d^{2} p_{\perp}^{\prime} \frac{h_{\;^1D_{s1}}^{\prime}}{x_{1} x_{2}\left(M^{\prime 2}
-M_{0}^{\prime 2}\right)}\left(\frac{m_{1}^{\prime}-m_{2}}{w_{\;^1D_{s1}}^{\prime}} p_{\perp}^{\prime 2}\right),
\en
where $m_{1}^{\prime}$ and $m_{2}$ represent the constituent quarks of the states $D^*_0, \ ^3D_{s1}$ and $^1D_{s1}$. The decay constants can be obtained through experimental measurements for the purely leptonic decays or theorical calculations. The explicit forms of $h'_{M}$  are given by \cite{Cheng:2003sm}
\be
h_{D^*_0}^{\prime} &=&\sqrt{\frac{2}{3}} h_{\;^3D_{s1}}^{\prime}=\left(M^{\prime 2}-M_{0}^{\prime 2}\right) \sqrt{\frac{x_{1} x_{2}}{N_{c}}} \frac{1}{\sqrt{2} \widetilde{M}_{0}^{\prime}}
\frac{\widetilde{M}_{0}^{\prime 2}}{2 \sqrt{3} M_{0}^{\prime}} \varphi_{p}^{\prime},\label{hs3a}\\
h_{\;^1D_{s1}}^{\prime} &=&\left(M^{2 \prime}-M_{0}^{\prime 2}\right) \sqrt{\frac{x_{1} x_{2}}{N_{c}}} \frac{1}{\sqrt{2} \widetilde{M}_{0}^{\prime}} \varphi_{p}^{\prime}.
\label{h1a}\en
\subsection{Form factors}
For the general $B\rightarrow M$ transitions with $M$ being a scalar or axial-vector meson \cite{Cheng:2003sm}, the decay amplitude at the lowest order is 
\be
\mathcal{M}^{B M}=-i^{3} \frac{N_{c}}{(2 \pi)^{4}} \int d^{4} p_{1}^{\prime} \frac{H_{B}^{\prime}\left(H_{M}^{\prime \prime}\right)}
{N_{1}^{\prime} N_{1}^{\prime \prime} N_{2}} S_{\mu}^{B M},
\en
where $N_{1}^{\prime(\prime \prime)}=p_{1}^{\prime(\prime \prime) 2}-m_{1}^{\prime (\prime\prime) 2}, N_{2}=p_{2}^{2}-m_{2}^{2} $ arise
from the quark propagators. For our considered transitions $B\rightarrow D^*_0$ and $B\to \ ^1D_{s1},\ ^3D_{s1}$,  
the traces $S_{\mu}^{BD^*_0}, S_{\mu \nu}^{B\;^1D_{s1}}$ and $S_{\mu \nu}^{B\;^3D_{s1}}$ can be directly obtained by using the Lorentz contraction as follows
\begin{footnotesize}
\begin{eqnarray}
S_{\mu}^{B D^*_0} &=& Tr\left[\left(\not p_{1}^{\prime \prime}+m_{1}^{\prime \prime}\right) \gamma_{\mu} \gamma_{5}\left(\not p_{1}^{\prime}
+m_{1}^{\prime}\right) \gamma_{5}\left(-\not p_{2}+m_{2}\right)\right],\\
S_{\mu \nu}^{B\;^1D_{s1}} &=&\left(S_{V}^{B \;^1D_{s1}}-S_{A}^{B \;^1D_{s1}}\right)_{\mu \nu} \non
&=&\operatorname{Tr}\left[\left(-\frac{1}{W_{\;^1D_{s1}}^{\prime \prime}}\left(p_{1}^{\prime \prime}-p_{2}\right)_{\nu}\right) \gamma_{5}\left(\not p_{1}^{\prime \prime}
+m_{1}^{\prime \prime}\right)\left(\gamma_{\mu}-\gamma_{\mu} \gamma_{5}\right)\left(\not p_{1}^{\prime}+m_{1}^{\prime}\right) \gamma_{5}\left(-\not p_{2}+m_{2}\right)\right],\;\;\;\label{btoa1}\\
S_{\mu \nu}^{B \;^3D_{s1}} &=&\left(S_{V}^{B \;^3D_{s1}}-S_{A}^{B \;^3D_{s1}}\right)_{\mu \nu} \non
&=&\operatorname{Tr}\left[\left(\gamma_{\nu}-\frac{1}{W_{\;^3D_{s1}}^{\prime \prime}}\left(p_{1}^{\prime \prime}-p_{2}\right)_{\nu}\right) \gamma_{5}\left(\not p_{1}^{\prime \prime}
+m_{1}^{\prime \prime}\right)\left(\gamma_{\mu}-\gamma_{\mu} \gamma_{5}\right)\left(\not p_{1}^{\prime}+m_{1}^{\prime}\right) \gamma_{5}\left(-\not p_{2}+m_{2}\right)\right].\;\;\;\label{btoa3}
\end{eqnarray}
\end{footnotesize}

To calculate the amplitudes for the transition form factors,
we need the Feynman rules for the meson-quark-antiquark vertices ($i\Gamma'_M$), which are listed as
\be
i \Gamma_{D^*_0}^{\prime}&=&-i H_{D^*_0}^{\prime},\\
i \Gamma_{\;^3D_{s1}}^{\prime}&=&-i H_{\;^3D_{s1}}^{\prime}\left[\gamma_{\mu}+\frac{1}{W_{\;^3D_{s1}}^{\prime}}\left(p_{1}^{\prime}-p_{2}\right)_{\mu}\right] \gamma_{5},\\
i \Gamma_{\;^1D_{s1}}^{\prime}&=&-i H_{\;^1D_{s1}}^{\prime}\left[\frac{1}{W_{\;^1D_{s1}}^{\prime}}\left(p_{1}^{\prime}-p_{2}\right)_{\mu}\right] \gamma_{5}.
\en
In practice, we employ the light-front decomposition of the Feynman loop momentum and integrate out the minus component using the contour method. Then additional spurious contributions being proportional to the light-like four-vector $\tilde{\omega}=(0,2,\bf{0}_\perp)$ will appear. While they can be eliminated by including the zero-mode contributions in a proper way. If the covariant vertex functions are not exhibit singularity during integration, the transition amplitudes will capture the singularities in the antiquark propagators. The specific rules for the $p^-$ integration have been derived in Refs. \cite{Jaus:1999zv,Cheng:2003sm}, and the relevant ones are summarized in Appendix A. The integration then leads to
\be
N_{1}^{\prime(\prime \prime)} &\rightarrow& \hat{N}_{1}^{\prime(\prime \prime)}=x_{1}\left(M^{\prime(\prime \prime 2}-M_{0}^{\prime(\prime \prime) 2}\right),\non
H_{B}^{\prime} &\rightarrow& h_{B}^{\prime},H_{M}^{\prime\prime} \rightarrow h_{M}^{\prime \prime},\non
W_{M}^{\prime \prime} &\rightarrow& w_{M}^{\prime \prime}\;\;\; (\text{ for the $^3D_{s1}$ and $^1D_{s1}$ states}), \non
\int \frac{d^{4} p_{1}^{\prime}}{N_{1}^{\prime} N_{1}^{\prime \prime} N_{2}} H_{B}^{\prime} H_{M}^{\prime \prime} S^{BM} & \rightarrow&-i \pi \int \frac{d x_{2} d^{2}
	p_{\perp}^{\prime}}{x_{2} \hat{N}_{1}^{\prime} \hat{N}_{1}^{\prime \prime}} h_{B}^{\prime} h_{M}^{\prime \prime} \hat{S}^{BM},
\en
where
\be
 w_{^3D_{s1}}^{\prime\prime}=\frac{\widetilde{M}_{0}^{\prime\prime 2}}{m_{1}^{\prime\prime}-m_{2}}, \quad w_{^1D_{s1}}^{\prime\prime}=2, \quad M_{0}^{\prime \prime 2}=\frac{p_{\perp}^{\prime \prime 2}+m_{1}^{\prime \prime 2}}{x_{1}}+\frac{p_{\perp}^{\prime \prime 2}+m_{2}^{2}}{x_{2}},
\label{vertex}
\en
with $p''_\perp=p'_\perp-x_2q_\perp$ and $\widetilde{M}_{0}^{\prime\prime}=\sqrt{M_{0}^{\prime\prime 2}-\left(m_{1}^{\prime\prime}-m_{2}\right)^{2}}$. The explicit forms of $h^{\prime\prime}_{M}$ have been given in Eq. (\ref{hs3a}) and Eq. (\ref{h1a}).

Using  Eq. (\ref{hs3a})-Eq. (\ref{vertex}) and taking the integration rules given in Refs. \cite{Jaus:1999zv,Cheng:2003sm}, the form factors $F^{BD^*_0}_1(q^2), F^{BD^*_0}_0(q^2)$ and
$A^{B D_{s1}}(q^2)$, $V_0^{B D_{s1}}(q^2)$, $V_1^{B D_{s1}}(q^2)$, $V_2^{B D_{s1}}(q^2)$
can be obtained directly, which are listed in Appendix C.
\subsection{Vertex Corrections and The Hard Spectator Function}
Within the framework of QCD factorization \cite{BBNS}, the short-distance nonfactorizable corrections including the vertex corrections
and hard spectator interactions are considered.
The modifications of the Wilson coefficients $a_{1,2}$ from the vertex corrections are given as
\be
a_{i}(\mu) \to a_{i}(\mu)+\frac{\alpha_{s}(\mu)}{4 \pi} C_{F} \frac{C_{i}(\mu)}{N_{c}} V_{i}(M_2), \;\;i=1,2,
\en
with $M_2$ being the meson emitted from the weak vertex. The vertex functions $V_{1,2}(M_2)$ are written as \cite{BBNS}
\be
V_{1,2}(M_2)=12\ln\displaystyle{\frac{m_b}{\mu}}-18
+\frac{2\sqrt{2N_c}}{f_{M_2}}\int_0^1 dx\, \Phi_{M_2}(x)\, g(x),
\en
where $f_{M_2}$ and $\Phi_{M_2}(x)$ are the decay constant and the twist-2 meson
distribution amplitude of the meson $M_2$, respectively. The hard kernel $g(x)$ is 
\begin{eqnarray}
g(x) &=& 3\left( \frac{1-2x}{1-x}\ln{x} -i\,\pi \right)\nonumber\\
& & +\left[ 2\,{\rm Li}_2(x)-\ln^2 x +\frac{2\ln
x}{1-x}-(3+2i\,\pi)\ln x - (x\leftrightarrow 1-x) \right].
\end{eqnarray}
The modifications of the Wilson coefficients $a_{1,2}$ from the hard spectator-scattering corrections arising from a hard gluon exchange between the emitted meson and the spectator quark are written as
\be
a_1(\mu)\to a_1(\mu)+
 \,{C_F\pi\alpha_sC_{2}\over
	N_c^2}H_1(M_1M_2),\;\;\; a_2(\mu)\to a_2(\mu)+
	\,{C_F\pi\alpha_sC_{1}\over
	N_c^2}H_2(M_1M_2),
\en
where the hard spectator functions $H_i(i=1,2)$ are defined as \cite{Cheng:2007st}
%\begin{small}
\be
 H_{i}\left(M_{1} M_{2}\right)=\frac{-f_B f_{M_1}}{{D(M_1M_2)}} \int_{0}^{1} \frac{d \rho}{\rho} \Phi_{B}(\rho) \int_{0}^{1} \frac{d \xi}{\bar{\xi}} \Phi_{M_{2}}(\xi) \int_{0}^{1} \frac{d \eta}{\bar{\eta}}\left[\pm\Phi_{M_{1}}(\eta)+r_{\chi}^{M_{1}} \frac{\bar{\xi}}{\xi} \Phi^P_{M_{1}}(\eta)\right],\;\;
\en
%\end{small}
with $\bar\xi= 1-\xi$ and $\bar\eta= 1-\eta$.
$\Phi_{M_1}$ and $\Phi^P_{M_1}$ are the twist-2 and twist-3 LCDAs of the meson $M_1$.
The definations of  $D(M_1M_2)$ and $r_{\chi}^{M_{1}}$ can be found in Ref. \cite{Cheng:2007st}.
\section{Numerical results and discussions} \label{numer}
\subsection{Transition Form Factors}
\begin{table}[H]
\caption{The values of the input parameters Refs. \cite{BN,pdg22,Cheng:2003id,Becirevic:1998ua,Li:2009wq}. }.
\label{tab:constant}
\begin{tabular*}{16.5cm}{@{\extracolsep{\fill}}l|ccccccc}
  \hline\hline
\textbf{Mass(\text{GeV})} &$m_{b}=4.8$
&$m_{c}=1.4$&$m_{s}=0.37$&$m_{u,d}=0.25$&$m_{B}=5.279$   \\[1ex]
&$m_{\pi}=0.140$&$m_{K}=0.494$&$m_{\rho}=0.775$&$m_{K^{\ast}}=0.892$& $m_{B_{s}}=5.367$\\[1ex]
& $m_{D_{0}^{\ast}}=2.343$& $m_{D_{s0}^{\ast}}=2.317$  & $m_{D}=1.86966 $& $m_{D_{s}}=1.96835 $ \\[1ex]
& $m_{D_{s1}}=2.460$& $m_{D_{s1}^{'}}=2.536$  & $m_{D_{s}^{\ast}}=2.1122$  & $m_{D^{\ast}}=2.010 $ \\[1ex]
\hline
\end{tabular*}

\begin{tabular*}{16.5cm}{@{\extracolsep{\fill}}l|ccccc}
\hline
\textbf{ Decay constants(\text{GeV})} & $f_{\pi}=0.13$ &$f_{\rho}=0.209\pm0.002$ \\[1ex]
 & $f_{K}=0.16$&$f_{K^{\ast}}=0.217\pm0.005$\\[1ex]
 & $f_{B}=0.19\pm0.02$ & $f_{B_{s}}=0.231\pm0.015$\\[1ex]
 &$f_D=0.2046\pm0.005$&$f_{D_s}=0.2575\pm0.0046$\\[1ex]
 &$f_{D^*}=0.245\pm0.02^{+0.003}_{-0.002}$&$f_{D^*_s}=0.272\pm0.016^{+0.003}_{-0.020}$\\[1ex]
 &$f_{D_{s1}}=0.145\pm0.011$&$f_{D^{\prime}_{s1}}=0.032\pm0.006$\\[1ex]
&$f_{D_{s1}^{3/2}}=0.05$&$f_{D_{s1}^{1/2}}=0.145$\\[1ex]
&$f_{^3D_{s1}}=-0.121$&$f_{^1D_{s1}}=0.038$\\[1ex]
& $f_{D_{0}^{\ast}}=0.103\pm0.021$ & $f_{D_{s0}^{\ast}}=0.067\pm0.013$\\[1ex]
\hline\hline
\end{tabular*}
\label{constants}
\end{table}
The input parameters, such as the masses of the initial and final mesons, the decay constants, are listed in Table \ref{tab:constant}. The decay constants of the axial mesons $D_{s1}$ and $D^\prime_{s1}$
can be obtained from $f_{D_{s 1}^{1 / 2}}$ and $f_{D_{s 1}^{3 / 2}}$ through the mixing between $D_{s1}^{1/2}$ and $D_{s1}^{3/2}$ 
\be
f_{D_{s 1}}&=&f_{D_{s 1}^{1 / 2}} \cos \theta_{s}+f_{D_{s 1}^{3 / 2}} \sin \theta_{s},\nonumber\\
f_{D_{s 1}^{\prime}}&=&-f_{D_{s 1}^{1 / 2}} \sin \theta_{s}+f_{D_{s 1}^{3 / 2}} \cos \theta_{s}.
\en
Using these decay constants and the masses of the
constituent quarks and mesons given in Table \ref{tab:constant}, we can obtain the values of the shape parameters $\beta'$ for
our considered mesons, which are listed in Table \ref{beta}.
\begin{table}[H]
	\caption{The shape parameters $\beta'$ (in units of GeV) in the Gaussian-type light-front wave functions defined in Eq. (\ref{betap}), and the uncertainties are from the decay constants.}
	\begin{center}
		\scalebox{0.9}{
			\begin{tabular}{ccccc}
				\hline\hline
				$\beta^{'}_{\pi}$&$\beta^{'}_{K}$&$\beta^{'}_{\rho}$&$\beta^{'}_{K^*}$&$\beta^{'}_{D}$\\
				0.317&0.37&$0.261^{+0.001}_{-0.002}$&$0.279\pm0.004$&$0.464^{+0.011}_{-0.014}$\\
				\hline
				$\beta^{'}_{B}$&$\beta^{'}_{B_{s}}$&$\beta^{'}_{D_{s}}$&$\beta^{'}_{D^*}$&$\beta^{'}_{D^*_{s}}$\\
				$0.555^{+0.048}_{-0.048}$&$0.628^{+0.035}_{-0.034}$&$0.497^{+0.032}_{-0.028}$&$0.409^{+0.021}_{-0.022}$&$0.438^{+0.016}_{-0.027}$\\
				\hline
				$\beta^{'}_{D_{0}^{\ast}}$&$\beta^{'}_{D_{s0}^{\ast}}$&$\beta^{'}_{^3D_{s1}}$&$\beta^{'}_{^1D_{s1}^{'}}$&$$\\
				$0.373^{+0.063}_{-0.059}$&$0.325^{+0.043}_{-0.043}$&$0.342^{+0.030}_{-0.034}$&$0.342^{+0.039}_{-0.039}$&$$\\
				\hline\hline
			\end{tabular}\label{beta}
		}
	\end{center}
\end{table}

It is noticed that all the computations are conducted within the $q^+=0$  reference frame, where the form factors can only be obtained at spacelike momentum transfers $q^2=-q^2_{\bot}\leq0$. It is necessary to know the form factors in the timelike region for the physical decay processes. Here, we utilize the following double-pole approximation to parameterize the form factors in the spacelike region and then extend to the timelike region,
\be
F\left(q^{2}\right)=\frac{F(0)}{1-a q^{2} / m^{2}+b q^{4} / m^{4}},\label{F}
\en
where $m$ represents the initial meson mass and $F(q^{2})$ denotes the different form factors.
The values of $a$ and $b$ can be obtained by performing a 3-parameter fit to the form factors in the range $-15 \text{GeV}^2\leq q^2\leq0$, which are collected in Tables {\ref{form factor1} and \ref{form factor2}}. The uncertainties arise from the decay constants of the initial $B_{(s)}$ meson and the final state mesons.

In Table \ref{form factor1}, we list the form factors of the transitions $B_{(s)}\to D^*_0, D^*_{s0}, D_{(s)}, D^*_{(s)}$. One can find that
the form factors of the transitions $B_{(s)}\to D^*_0, D^*_{s0}$ are much smaller. This conclusion is also supported by other works, for example, the form factor of the transtion $B\to D^*_0$ was obtained as 0.24 and 0.18 within the CLFQM \cite{Cheng:2003sm} and the 2nd version of the Isgur-Scora-Grinstein-Wise (ISGW2) approach \cite{Cheng:2003id}, respectively. Furthermore, our result for the form factor of the transtion $B_s\to D^*_{s0}$ is
consistent with 0.20 gvien in the ISGW2 model \cite{Cheng:2003id}, while smaller than 0.40 given by the QCD sum rules (QCDSR) approach \cite{T.M.}. As to the form factors of the transitions $B_{(s)}\to D_{(s)}, D^*_{(s)}, \pi (\rho), K (K^*)$,  they have been searched by many appoaches, such as the Melikhov-Stech (MS) model \cite{Melikhov:2000yu}, the relaticistic quark model (RQM) \cite{Kramer2}, the BSW model \cite{bsw,Kramer:1992xr}, the Bethe-Salpeter (BS) equation \cite{Chen:2011ut}, the QCDSR \cite{Blasi:1993fi} and the light cone sum rules (LSCR) approach \cite{Ball:1998tj}. Cosidering the need for the latter branching ratio calcuations, we also give them in Table \ref{form} with other theoretical results for comparison. Obviously, our predictions are consistent well with these theoretical results.  
\begin{table}[H]
	\caption{Form factors of the transitions $B_{(s)}\to D^*_0, D^*_{s0}, D_{(s)}, D^*_{(s)}$ in the CLFQM.
		The uncertainties are from the decay constants of $B_{(s)}$ and final state mesons.}
	\begin{center}
		\scalebox{1.0}{
			\begin{tabular}{ccccc}
				\hline\hline
				& $F_{i}(q^{2}=0)$&$F_{i}(q^{2}_{max})$&a&b\\
				\hline\hline
				$F_{1}^{BD^{\ast}_{0}}$&$0.25^{+0.03+0.05}_{-0.02-0.05}$&$0.30^{+0.03+0.06}_{-0.03-0.07}$&$0.70^{+0.04+0.03}_{-0.05-0.11}$&$0.65^{+0.08+0.03}_{-0.07-0.07}$\\
				$F_{0}^{BD^{\ast}_{0}}$&$0.25^{+0.03+0.05}_{-0.02-0.05}$&$0.22^{+0.02+0.04}_{-0.01-0.04}$&$-0.38^{+0.04+0.05}_{-0.04-0.02}$&$0.21^{+0.07+0.08}_{-0.07-0.08}$\\
				\hline
				$F_{1}^{B_s D^*_{s0}}$&$0.21^{+0.02+0.04}_{-0.01-0.04}$&$0.24^{+0.02+0.05}_{-0.01-0.05}$&$0.63^{+0.05+0.07}_{-0.06-0.12}$&$0.78^{+0.08+0.01}_{-0.09-0.04}$\\
				$F_{0}^{B_s D^*_{s0}}$&$0.21^{+0.02+0.04}_{-0.01-0.04}$&$0.18^{+0.02+0.03}_{-0.01-0.03}$&$-0.43^{+0.01+0.01}_{-0.00-0.02}$&$0.28^{+0.03+0.01}_{-0.06-0.04}$\\
				\hline
				$F_{1}^{ B D}$&$0.66^{+0.00+0.00}_{-0.01-0.01}$&$0.81^{+0.00+0.00}_{-0.01-0.02}$&$0.80^{+0.01+0.00}_{-0.01-0.01}$&$0.86^{+0.03+0.01}_{-0.03-0.01}$\\
				$F_{0}^{ B D}$&$0.66^{+0.00+0.00}_{-0.01-0.01}$&$0.70^{+0.02+0.00}_{-0.02-0.01}$&$0.46^{+0.01+0.01}_{-0.00-0.00}$&$0.78^{+0.10+0.03}_{-0.10-0.02}$\\
				\hline
				$F_{1}^{ B_s D_{s}}$&$0.65^{+0.00+0.01}_{-0.00-0.02}$&$0.79^{+0.00+0.01}_{-0.00-0.03}$&$0.84^{+0.01+0.01}_{-0.01-0.02}$&$1.02^{+0.03+0.04}_{-0.04-0.05}$\\
				$F_{0}^{ B_s D_{s}}$&$0.65^{+0.00+0.01}_{-0.00-0.02}$&$0.68^{+0.01+0.01}_{-0.01-0.03}$&$0.50^{+0.01+0.01}_{-0.01-0.02}$&$0.99^{+0.07+0.09}_{-0.07-0.10}$\\
				\hline
				$V^{B D^{\ast}}$&$0.73^{+0.01+0.01}_{-0.01-0.02}$&$ 0.89^{+0.01+0.01}_{-0.02-0.03}$&$0.82^{+0.01+0.00}_{-0.01-0.01}$&$0.91^{+0.05+0.02}_{-0.05-0.02}$      \\
				$A^{B D^{\ast}}_{0}$&$0.67^{+0.00+0.02}_{-0.01-0.02}$&$ 0.70^{+0.00+0.02}_{-0.02-0.03}$&$0.16^{+0.02+0.00}_{-0.03-0.01}$& $0.15^{+0.00+0.00}_{-0.01-0.01}$      \\
				$A^{B D^{\ast}}_{1}$&$0.63^{+0.00+0.01}_{-0.00-0.01}$&$ 0.72^{+0.01+0.01}_{-0.00-0.01}$&$0.42^{+0.02+0.01}_{-0.02-0.01}$ & $0.22^{+0.03+0.02}_{-0.01-0.00}$        \\
				$A^{B D^{\ast}}_{2}$&$0.59^{+0.00+0.00}_{-0.01-0.01}$&$ 0.71^{+0.00+0.00}_{-0.01-0.01}$ &$0.75^{+0.01+0.01}_{-0.02-0.02}$ & $0.78^{+0.04+0.03}_{-0.04-0.03}$        \\
				\hline
				$V^{B_s D^{\ast}_{s}}$&$0.72^{+0.01+0.02}_{-0.00-0.02}$&$0.86 ^{+0.02+0.03}_{-0.00-0.02}$&$0.86^{+0.02+0.01}_{-0.02-0.00}$ & $1.10^{+0.05+0.04}_{-0.05-0.02}$      \\
				$A^{B_s D^{\ast}_{s}}_{0}$&$0.65^{+0.01+0.02}_{-0.00-0.03}$&$ 0.69^{+0.01+0.02}_{-0.01-0.04}$&$0.23^{+0.02+0.00}_{-0.03-0.01}$& $0.21^{+0.01+0.02}_{-0.01-0.01}$      \\
				$A^{B_s D^{\ast}_{s}}_{1}$&$0.62^{+0.00+0.01}_{-0.01-0.02}$&$ 0.72^{+0.00+0.00}_{-0.02-0.03}$&$0.48^{+0.01+0.00}_{-0.03-0.02}$ & $0.32^{+0.03+0.03}_{-0.02-0.01}$        \\
				$A^{B_s D^{\ast}_{s}}_{2}$&$0.57^{+0.00+0.00}_{-0.00-0.00}$&$0.68^{+0.00+0.00}_{-0.00-0.00} $ &$0.80^{+0.01+0.01}_{-0.02-0.02}$ & $0.95^{+0.04+0.06}_{-0.04-0.03}$        \\
				\hline\hline
			\end{tabular}\label{form factor1}
		}
	\end{center}
\end{table}
\begin{table}[H]
	\caption{Form factors of the transitions $B_{(s)}\to D_{(s)}, D^*_{(s)}, \pi (\rho), K (K^*)$ at $q^{2}= 0$ together with other theoretical results. }
	\begin{center}
		\scalebox{1.0}{
			\begin{tabular}{c|c|ccccc}
				\hline\hline
				Transitions  &References&\;\;$F_{0}(0)\;\;$&\;\;$V(0)$&\;\;$A_{0}(0)$&\;\;$A_{1}(0)$&\;\;$A_{2}(0)$\\
				\hline\hline
				$B\to D,D^* $&This work&$0.66$&$0.73$&$0.67$&$0.63$&$0.59$\\
				\hline
				$ $&\cite{Cheng:2003sm}&$0.67$&$0.75$&$0.64$&$0.63$&$0.62$\\
				$ $&\cite{Melikhov:2000yu}&$0.67$&$0.76$&$0.69$&$0.66$&$0.62$\\
				$ $&\cite{bsw}&$0.69$&$0.71$&$0.62$&$0.65$&$0.69$\\
				\hline
				$B_s \to D_s,D^*_s$&This work&$0.65$&$0.57$&$0.72$&$0.65$&$0.62$\\
				\hline
				$ $&\cite{Kramer2}&$0.74$&$0.95$&$0.67$&$0.70$&$0.75$\\
				$ $&\cite{Kramer:1992xr}&$0.61$&$0.64$&$-$&$0.56$&$0.59$\\
				$ $&\cite{Chen:2011ut} &$0.57$&$0.70$&$0.70$&$0.65$&$0.67$  \\
				$ $&\cite{Blasi:1993fi}&$0.70$&$0.63$&$-$&$0.62$&$0.75$ \\
			
				\hline
				$B \to \pi,\rho$&This work&$0.25$&$0.25$&$0.28$&$0.22$&$0.19$\\
				\hline
				$ $&\cite{Cheng:2003sm}&$0.25$&$0.27$&$0.28$&$0.22$&$0.20$\\
				$ $&\cite{Ball:1998tj}&$0.26$&$0.34$&$0.37$&$0.26$&$0.22$\\
				$ $&\cite{Melikhov:2000yu}&$0.29$&$0.31$&$0.29$&$0.26$&$0.24$\\
				$ $&\cite{bsw}&$0.33$&$0.33$&$0.28$&$0.28$&$0.28$\\
				\hline
				$B \to K,K^*$&This work&$0.35$&$0.30$&$0.33$&$0.26$&$0.24$\\
				\hline
				$ $&\cite{Cheng:2003sm}&$0.35$&$0.31$&$0.31$&$0.26$&$0.24$\\
				$ $&\cite{Ball:1998tj}&$0.34$&$0.46$&$0.47$&$0.34$&$0.28$\\
				$ $&\cite{Melikhov:2000yu}&$0.36$&$0.44$&$0.45$&$0.36$&$0.32$\\
				$ $&\cite{bsw}&$0.38$&$0.37$&$0.32$&$0.33$&$0.33$\\
				\hline\hline
			\end{tabular}\label{form}
		}
	\end{center}
\end{table}

\begin{figure}[H]
	\vspace{0.32cm}
	\centering
	\subfigure[]{\includegraphics[width=0.45\textwidth]{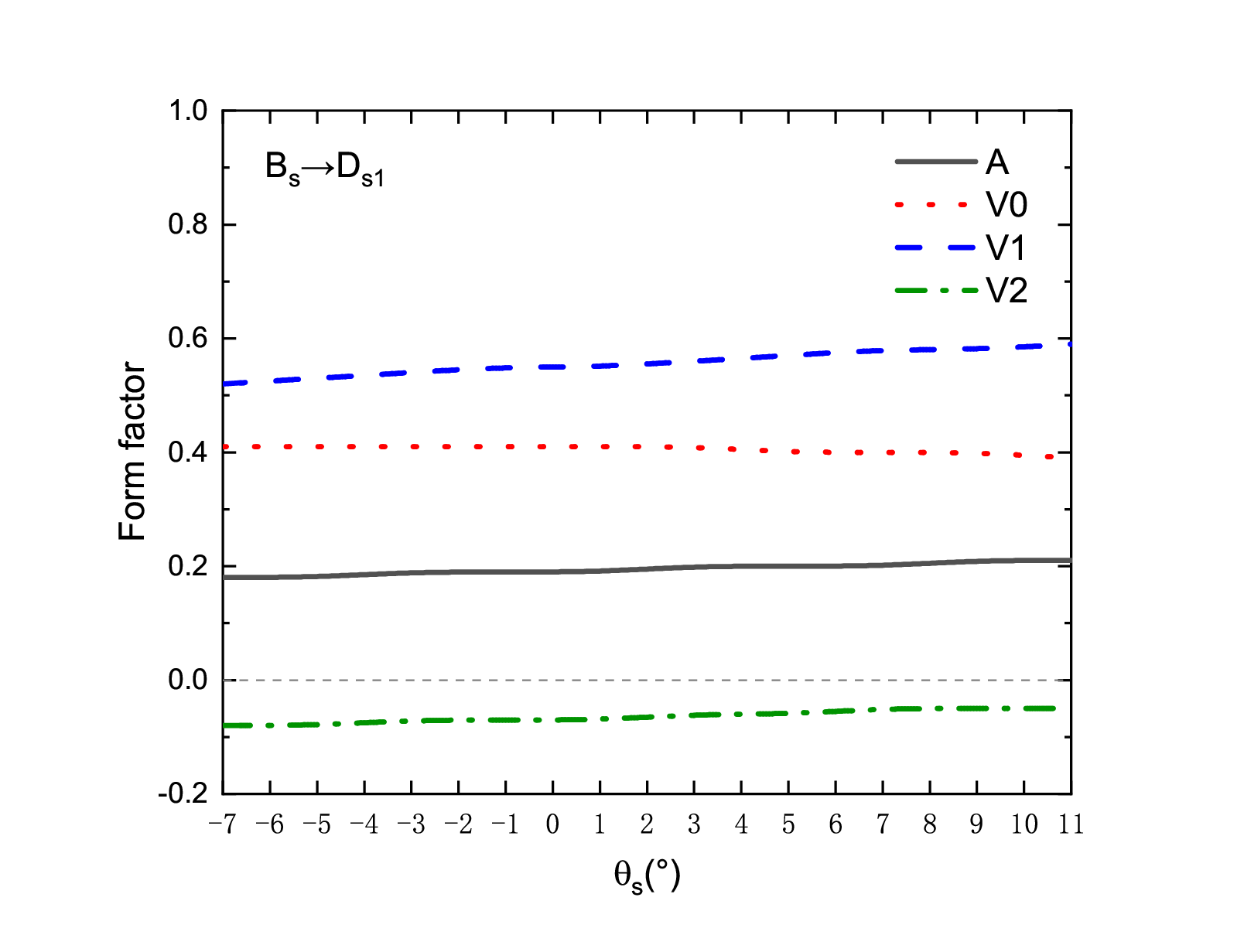}\quad}
	\subfigure[]{\includegraphics[width=0.45\textwidth]{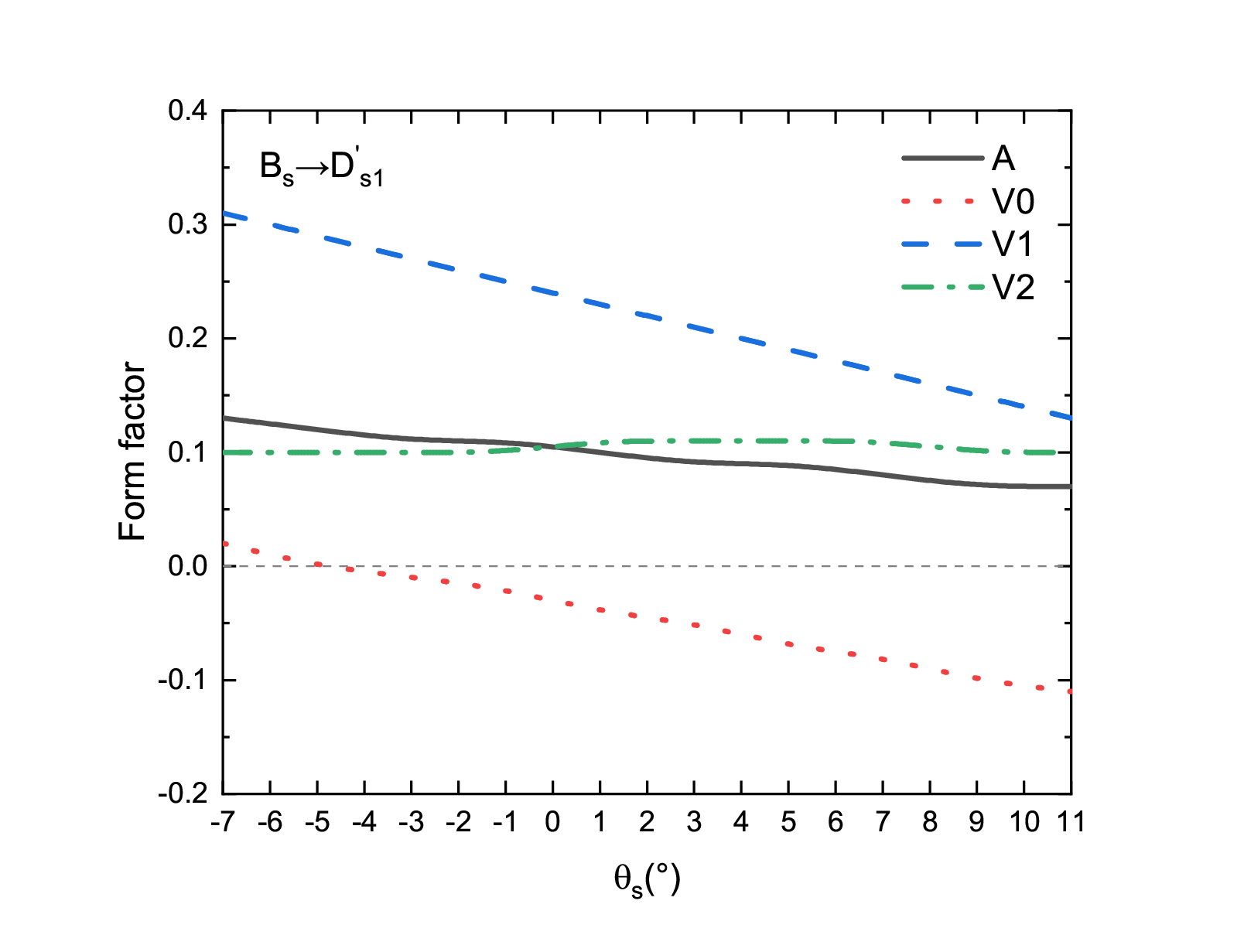}}\\
	\caption{The dependence of the form factors of the transitions $B_s\to D_{s1}$ and $ B_s\to D_{s1}^\prime$ on the mixing angle $\theta_s$. }\label{fig:T2}
\end{figure}

 In order to determine the physical form factors of the transitions $B_s\to D_{s1}, D_{s1}^\prime$, we need to know the mixing angle $\theta=\theta_s+35.3^\circ$ between $^1D_{s1}$ and $^3D_{s1}$ shown in Eq. (\ref{mixing3}). Here we take $\theta_s=7^\circ$, which was determined from the quark potential model \cite{Cheng:2003id}. The results are listed in Table \ref{form factor2}, where the uncertainties are from the decay constants of $B_{s}$ and the final states ($^3D_{s1}$, $^1D_{s1}$). In Figure \ref{fig:T2}, we check the dependence of the form factors of the transitions $B_s\to D_{s1}, D_{s1}^\prime$ on the mixing angle $\theta_s$. We find that 
the form factor $V_0$ of the transition  $B_s\to D_{s1}^\prime$ ($B_s\to  D_{s1}$)  is (not) sensitive to the mixing angle. It can be used to explain why the branching ratios of the decays associated with the transition $B_s\to D_{s1}^\prime$ ($B_s\to D_{s1}$) are (not) sensitive to the mixing angle, which will be discussed in the latter.
\begin{table}[H]
	\caption{The form factors of the transitions $B_s\to D_{s1}$ and $B_s\to D^{'}_{s1}$ in the CLFQM.
		The uncertainties are from the decay constants of $B_{(s)}$ and final states.}
	\begin{center}
		\scalebox{0.85}{
			\begin{tabular}{ccccc}
				\hline\hline
				& $F_{i}(q^{2}=0)$&$F_{i}(q^{2}_{max})$&a&b\\
			\hline
			$A^{ B_s D_{s1}}$&$0.20^{+0.01+0.02+0.01}_{-0.01-0.02-0.00}$&$0.18^{+0.02+0.02+0.02}_{-0.01-0.02-0.00} $&$-0.27^{+0.06+0.03+0.07}_{-0.07-0.05-0.08}$ & $0.11^{+0.02+0.01+0.02}_{-0.02-0.01-0.03}$\\
			$V^{ B_s D_{s1}}_{0}$&$0.40^{+0.02+0.01+0.04}_{-0.02-0.00-0.04}$&$0.42^{+0.02+0.01+0.05}_{-0.02-0.00-0.05} $&$-0.17^{+0.02+0.02+0.03}_{-0.04-0.04-0.04}$ & $-0.02^{+0.01+0.00+0.00}_{-0.00-0.00-0.01}$\\
			$V^{ B_s D_{s1}}_{1}$&$0.58^{+0.01+0.02+0.02}_{-0.02-0.03-0.03}$&$0.57^{+0.01+0.02+0.02}_{-0.02-0.03-0.03} $&$-0.05^{+0.01+0.01+0.00}_{-0.01-0.01-0.00}$ & $0.02^{+0.00+0.01+0.01}_{-0.00-0.00-0.00}$\\
			$V^{ B_s D_{s1}}_{2}$&$-0.05^{+0.01+0.01+0.02}_{-0.00-0.00-0.01}$&$ -0.05^{+0.01+0.01+0.03}_{-0.00-0.01-0.02}$&$0.56^{+0.06+0.12+0.18}_{-0.06-0.15-0.20}$ & $2.50^{+0.25+1.11+1.25}_{-0.20-0.85-0.98}$\\
			\hline
			$A^{ B_s D^{'}_{s1}}$&$0.08^{+0.01+0.02+0.00}_{-0.01-0.02-0.01}$&$0.03^{+0.01+0.01+0.00}_{-0.02-0.03-0.00}$&$2.05^{+0.13+0.24+0.26}_{-0.10-0.26-0.24}$ & $5.57^{+0.25+0.34+0.36}_{-0.20-0.26-0.32}$\\
			$V^{B_s D^{'}_{s1}}_{0}$&$-0.08^{+0.01+0.01+0.04}_{-0.01-0.00-0.04}$&$-0.05^{+0.02+0.02+0.04}_{-0.01-0.00-0.04}$&$1.24^{+0.05+0.14+0.18}_{-0.06-0.17-0.18}$ & $0.74^{+0.02+0.16+0.14}_{-0.02-0.10-0.14}$\\
			$V^{ B_s D^{'}_{s1}}_{1}$&$0.17^{+0.02+0.03+0.02}_{-0.03-0.02-0.02}$&$0.15^{+0.01+0.02+0.01}_{-0.03-0.02-0.02}$&$-0.52^{+0.06+0.02+0.05}_{-0.05-0.02-0.05}$ & $0.36^{+0.01+0.02+0.02}_{-0.00-0.03-0.07}$\\
			$V^{ B_s D^{'}_{s1}}_{2}$&$0.11^{+0.01+0.00+0.01}_{-0.02-0.01-0.02}$&$0.10^{+0.01+0.00+0.01}_{-0.02-0.01-0.02}$&$0.25^{+0.06+0.00+0.06}_{-0.07-0.00-0.07}$ & $-0.07^{+0.03+0.00+0.01}_{-0.04-0.01-0.03}$\\
				\hline
			
				\hline\hline
			\end{tabular}\label{form factor2}
		}
	\end{center}
\end{table}

\begin{figure}[H]
	\vspace{0.32cm}
	\centering
	\subfigure[]{\includegraphics[width=0.40\textwidth]{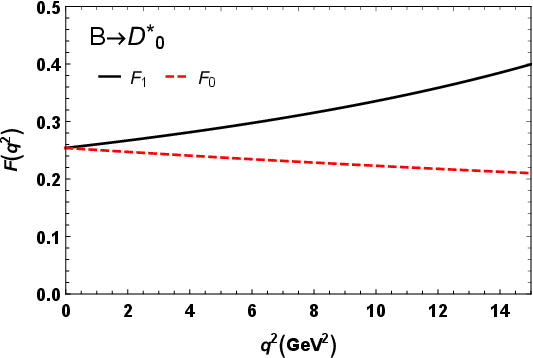}\quad}
	\subfigure[]{\includegraphics[width=0.40\textwidth]{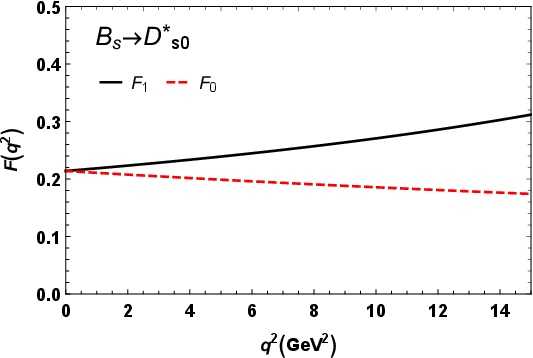}}\\
	\subfigure[]{\includegraphics[width=0.425\textwidth]{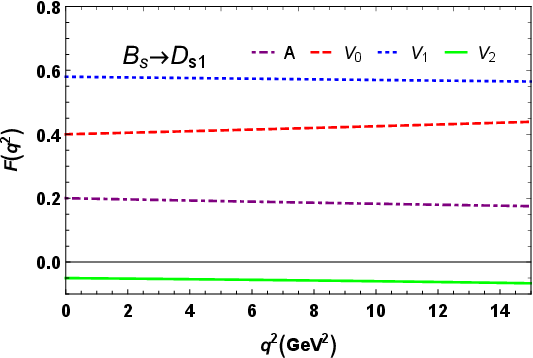}}
	\subfigure[]{\includegraphics[width=0.425\textwidth]{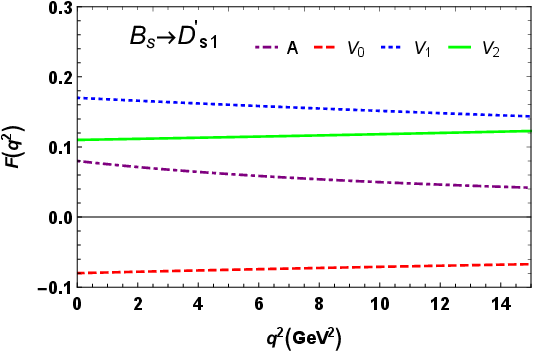}}\\
	\caption{The $q^2$-dependence of the form factors of the transitions $B_{(s)}\to D_{0}^{\ast}, D_{s0}^{\ast}, D_{s1}, D_{s1}^\prime$. }\label{fig:T1}
\end{figure}

In Figure \ref{fig:T1}, we plot the $q^2$-dependence of the form factors of the transitions  $B_{(s)}\to D^*_{0}, D^*_{s0}, D_{s1}, D_{s1}^\prime$. There exists the similar $q^2$-dependence of the form factors 
$F_{0,1}(q^2)$ between the transitions $B\to D_{0}^{\ast}$ and $B_{s}\to D_{s0}^{\ast}$. It is consistent with our expectation. While it is very different in magnitude between the corresponding form factors for the transitions $B_{s}\to D_{s1}$ and $B_{s}\to D_{s1}^\prime$ .

\subsection{Branching ratios}
In addition to the parameters listed in Table \ref{tab:constant}, other inputs, such as the $B_{(s)}$ meson lifetime $\tau_{B_{(s)}}$, the Wilson coefficients $a_{1},a_{2}$ and the CabibboKobayashi-Maskawa (CKM) matrix elements, are listed as \cite{pdg22,Beneke:2001ev}
\be
\tau_{B^{\pm}}&=&(1.519\pm0.004)\times10^{-12}s,\tau_{B_{s}}=(1.520\pm0.005)\times10^{-12}s\\
\tau_{B_{0}}&=&(1.638\pm0.004)\times10^{-12}s, a_1=1.018,a_2=0.17, \\
V_{cd}&=&0.221\pm0.004,V_{cb}=(40.8\pm1.4)\times10^{-3}, V_{cs}=0.975\pm0.006 \\
V_{us}&=&0.2243\pm0.0008, V_{ud}=0.97373\pm0.00031 .
\en
Firstly, we consider the branching ratios of the decays $ B_{(s)}\to D^{\ast}_{(s)0}M$ with $M$ being a light pseudoscalar (vector) meson $P(V)$ or a charmed meson $(D^{(*)},D^{(*)}_s)$  , which can be calculated through the formula
\be
\mathcal{B}r(B_{(s)}\to D^*_{(s)0} M)=\frac{\tau_{B_{(s)}}}{\hbar}\Gamma(B_{(s)}\to D^*_{(s)0} M),\label{branch}
\en
where the decay width $\Gamma( B_{(s)}\to D^{\ast}_{0}(D^{\ast}_{s0})M)$ for each channel is given as following
\be
\Gamma\left(B^0 \to D^{\ast-}_{0} P(V)\right) & =
& \frac{\left|G_{F} V_{c b} V_{u q}^{*} a_{1} f_{P(V)}
m_{B}^{2} F_{0}^{BD^{\ast}_{0}}(m^2_{P(V)})\right|^{2}}{32 \pi
m_{B}}\left(1-r_{D^{\ast}_{0}}^{2}\right),\\
\Gamma\left(B^0 \to D^{\ast-}_{0} D^{(*)+}\right) & =
& \frac{\left|G_{F} V_{c b} V_{c d}^{*} a_{1} f_{D^{(*)}}
m_{B}^{2} F_{0}^{BD^{\ast}_{0}}(m^2_{D^{(*)}})\right|^{2}}{32 \pi
m_{B}}\left(1-r_{D^{\ast}_{0}}^{2}-r_{D^{(*)}}^{2}\right), \label{Ds0D}
\en
where $P(V)$ refers to $\pi, K$ ($\rho, K^*$) and the subscript $q=d (s)$. $\Gamma\left(B^0 \to D^{\ast-}_{0} D^{(*)+}_s\right)$ can be obtained by replacing $V_{c d}, D^{(*)}$ with $V_{c s}$, $D^{(*)}_s$ in Eq. (\ref{Ds0D}). There exists similar expressions for the cases with $B^0, D^{\ast-}_{0}$ replaced by $B^0_s, D^{\ast-}_{s0}$ in the upper decays. While for the charged $B$ decays $B^+\to \bar D^{*0}_0 P(V)$ with $P(V)$ being $\pi^+, K^+ (\rho^+, K^{\ast+})$, the corresponding decay widths should be written as
\be
\Gamma\left(B^+ \to \bar{D}^{*0}_0\pi^+(K^+) \right)  &=&
 \frac{\left(G_{F} V^{*}_{c b} V_{u q}m_{B}^{2}\right)^{2}\left(1-r_{ D^*_0}^{2}\right)}{32 \pi m_{B}}\non &&
 \times|a_{1} f_{\pi(K)} F_{0}^{B D^*_0}(m_{\pi(K)}^2)+ a_{2} f_{D^*_0}
F_{0}^{B \pi(K)}(m_{D^*_0}^2)|^2, \label{chi1p}\\
\Gamma\left(B^+ \to \bar{D}^{*0}_0\rho^+ (K^{\ast+})\right)  &=&
 \frac{\left(G_{F} V^{*}_{c b} V_{u q}m_{B}^{2}\right)^{2}\left(1-r_{D^*_0}^{2}\right)}{32 \pi m_{B}}\non &&
 \times|a_{1} f_{\rho (K^{\ast})}F_{0}^{B D^*_0}(m_{\rho (K^{\ast})}^2)+ a_{2} f_{D^*_0}
 A_{0}^{B \rho (K^{\ast})}(m_{D^*_0}^2)|^2.
 \en
 Secondly, the decay widths of the decays $B_{s} \to D_{s1}^{(\prime)} M$ are defined as
 \be
 \Gamma\left(B_{s} \to D_{s1}^{(\prime)} \pi(K)\right) & =
 & \frac{\left|G_{F} V_{c b} V_{u q}^{*} a_{1} f_{\pi(K)}
 	m_{B_{s}}^{2} V_{0}^{B_{s} D_{s1}^{(\prime)}}(m^2_{\pi(K)})\right|^{2}}{32 \pi
 	m_{B_{s}}}\left(1-r_{D_{s1}^{(\prime)}}^{2}\right),\\
 	\Gamma\left(B_{s} \to D_{s1}^{(\prime)} D_{(s)}\right) & =
 	& \frac{\left|G_{F} V_{c b} V_{c q}^{*} a_{1} f_{D_{(s)}}
 		m_{B_{s}}^{2} V_{0}^{B_{s} D_{s1}^{(\prime)}}(m^2_{D_{(s)}})\right|^{2}}{32 \pi
 		m_{B_{s}}}\left(1-r_{D_{s1}^{(\prime)}}^{2}-r_{D_{(s)}}^{2}\right), \label{Ds1D}\\
\Gamma\left(B_{s}\to D_{s1}^{(\prime)} V\right) & = & \frac{|\vec{p}|}{8 \pi m_{B_{s}}^{2}}\left(
|\mathcal{A}_{L}(B_{s}\to D_{s1}^{(\prime)}V)|^2+2\left|\mathcal{A}_{N}\left(B_{s}\to D_{s1}^{(\prime)} V\right)\right|^{2}\right.\non &&\left.+2\left|\mathcal{A}_{T}\left(B_{s}\to D_{s1}^{(\prime)} V\right)\right|^{2}\right),\label{Ds1Dstar}
\en
where $V$ represents a vector meson, such as $\rho,K^*, D^*_{(s)}$, and the summation
of the three polarizations for the decays $B_s\to D_{s1}^{(\prime)}V$ are performed.
The three-momentum $\vec{p}$ is defined as
\be
|\vec{p}|=\frac{\sqrt{\left(m_{B_{s}}^{2}-\left(m_{ D_{s1}^{(\prime)}}+m_{V}\right)^{2}\right)\left(m_{B_{s}}^{2}-
		\left(m_{ D_{s1}^{(\prime)}}-m_{V}\right)^{2}\right)}}{2 m_{B_{s}}},
\en
and the three polarization amplitudes ${\cal A}_L,{\cal A}_N,$ and ${\cal A}_T$ are given as
\begin{footnotesize}
\be
i {\cal A}_L(B_{s}\to D_{s1}^{(\prime)} V )&=& \frac{ (-i)^3G_F}{\sqrt 2}V_{cb}V_{q_1q_2}^* a_1   f_{ V }m_{B_{s^{}}}^2\frac{ 1}{2 r_{D_{s1}^{(\prime)}}} \nonumber\\
&&\times \left[(1-r_{ V }^2-r_{D_{s1}^{(\prime)}}^2)(1-r_{D_{s1}^{(\prime)}})V_1^{B_s D_{s1}^{(\prime)}}(m_{ V }^2)-\frac{{\lambda(1,r_{ V }^2, r_{D_{s1}^{(\prime)}}^2)}}{1-r_{D_{s1}^{(\prime)}}}V_2^{B_{s}D_{s1}^{(\prime)}}(m_{ V }^2)\right],\nonumber\\
	i{\cal A}_N(B_{s}\to D_{s1}^{(\prime)} V )&=&
	\frac{ (-i)^3G_F}{\sqrt 2}
	V_{cb}V_{q_1q_2}^* a_1  f_{ V } m_{B_{s}}^2 r_{ V } (1-r_{D_{s1}^{(\prime)}})  V_1^{B_{s} D_{s1}^{(\prime)}}(m_{ V }^2),\nonumber\\
	i {\cal A}_T(B_{s} \to D_{s1}^{(\prime)}  V )&=& \frac{-i G_F}{\sqrt 2}
	V_{cb}V_{q_1q_2}^* a_1   f_{ V }m_{B_{s}}^2  r_{ V } \frac{\sqrt {\lambda(1, r_{ V }^2,r_{D_{s1}^{(\prime)}}^2)}}{1-r_{D_{s1}^{(\prime)}}} A^{B_{s} D_{s1}^{(\prime)}}(m_{ V }^2),\label{transverse}
\en
\end{footnotesize}
with $q_1=u,c$ and $q_2=d,s$, $\lambda\left(1, r_{V}^{2}, r_{D_{s1}}^{2}\right)=\left(1+r_{V}^{2}-r_{D_{s1}}^{2}\right)^{2}-4r_{V}^{2}$. As to the decay widths of the decays $B\to D_{s1}^{(\prime)}D$ and $B\to D_{s1}^{(\prime)}D^*$, they can be obtained from Eqs. (\ref{Ds1D}) and (\ref{Ds1Dstar}) with some simple replacements, respectively. Using the upper listed input parameters and the formulas given in Eq. (\ref{branch})-Eq. (\ref{transverse}), one can calculate the branching ratios for the considered decays shown in Tables \ref{D2300pik}-\ref{last}. 
\begin{table}[H]
\caption{The CLFQM predictions for the branching ratios of the decays $B\to D^*_0\pi(K)$. The label LO, VC and HSSC mean the inclusions of the leading order, the vertex corrections and the hard spectator-scattering corrections, respectively. NLO means the inclusions of these two kinds of corrections at the same time. 
The upper (lower) line is corresponding to $f_{D^{\ast}_{0}}=0.078 (0.103)$ GeV for each decay. The first and second uncertainties are from the decay constants of $B$ and $D^*_0$.}
\begin{center}
\scalebox{0.7}{
\begin{tabular}{|c |c |c|c |c |c |c |}
\hline\hline
&LO &VC&HSSC&NLO&Other predictions& Refs.\\
\hline\hline
$10^{-4}\times\mathcal{B} r(B^{+}\to \bar{D}^{\ast0}_{0}\pi^{+})$&$2.98^{+1.07+1.72}_{-0.83-1.44}$&$4.68^{+1.48+2.15}_{-1.09-1.90}$&$2.60^{+1.00+1.61}_{-0.77-1.33}$&$4.18^{+1.27+2.03}_{-1.02-1.77}$&$8.93$&PQCD\;\;\;\cite{Wang:2018fai}\\
&\;&\; &\; &\; &\;  $7.3$\; & CLFQM\;\;\;\cite{Cheng:2003sm} \\
&\;$5.45^{+1.24+2.53}_{-1.09-2.36}$&\; $8.49^{+1.57+3.16}_{-1.42-3.08}$&\;$4.71^{+1.15+2.35}_{-1.10-2.16}$ &\;$7.59^{+1.48+2.98}_{-1.32-2.88}$ &\; $7.7$\; &ISGW2\;\;\cite{Cheng:2003id} \\
&\;&\; &\; &\; &\;  $4.2$\; &\;ISGW\;\;\cite{Katoch:1995hr}\\
\hline
$10^{-4}\times\mathcal{B} r({B}^{0}\to D^{\ast-}_{0}\pi^{+})$&$3.68^{+1.03+1.82}_{-0.91-1.58}$&$3.83^{+0.95+1.89}_{-1.18-1.65}$&$3.59^{+1.11+1.77}_{-0.89-1.55}$&$3.75^{+1.16+1.85}_{-0.93-1.61}$&$4.28$&PQCD\;\;\;\cite{Wang:2018fai}\\
&\;$6.69^{+1.32+2.66}_{-1.18-2.56}$&\; $6.97^{+1.37+2.78}_{-1.23-2.67}$&\;$6.54^{+1.29+2.60}_{-1.15-2.51}$ &\;$6.81^{+1.34+2.71}_{-1.20-2.61}$ &\;  $2.6$\; &ISGW2\;\;\cite{Cheng:2003id}\\
&\;&\; &\; &\; &\;  $4.1$\; &\;ISGW\;\;\cite{Katoch:1995hr}\\
\hline
$10^{-5}\times\mathcal{B} r(B^{+}\to \bar{D}^{\ast0}_{0} K^{+})$&$2.25^{+0.90+1.32}_{-0.68-1.11}$&$3.74^{+1.07+1.70}_{-0.86-1.51}$&$1.94^{+0.76+1.23}_{-0.58-1.01}$&$3.20^{+0.98+1.57}_{-0.78-1.37}$&$6.96$&PQCD\;\;\;\cite{Wang:2018fai}\\
&\;$4.12^{+0.95+1.95}_{-0.83-1.81}$&\;$6.78^{+1.24+2.50}_{-1.12-2.43}$ &\;$3.57^{+0.88+1.81}_{-0.76-1.66}$ &\;$6.08^{+1.17+2.35}_{-1.05-2.27}$ &$$&$$\\
\hline
$10^{-5}\times\mathcal{B} r({B}^{0}\to D^{\ast-}_{0}K^{+})$&$2.91^{+0.90+1.44}_{-0.72-1.25}$&$3.03^{+0.75+1.50}_{-0.94-1.30}$&$2.82^{+0.87+1.39}_{-0.70-1.21}$&$2.94^{+0.91+1.45}_{-0.73-1.26}$&$3.57$&PQCD\;\;\;\cite{Wang:2018fai}\\
 &\;$5.29^{+1.04+2.11}_{-0.93-2.03}$&\; $5.52^{+1.09+2.20}_{-0.97-2.11}$&\;$5.13^{+1.01+2.04}_{-1.00-1.97}$&\;$5.35^{+1.05+2.13}_{-0.94-2.05}$&$$&$$\\
\hline\hline
\end{tabular}\label{D2300pik}}
\end{center}
\end{table}

From Table \ref{D2300pik}, one can find that the branching ratios of the decays $B^{+}\to \bar{D}^{\ast0}_{0}\pi^{+}$ and $B^{+}\to \bar{D}^{\ast0}_{0}K^{+}$ are sensitive to the vertex corrections. These two channels are contributed by two kinds of Feynman diagrams, one is associated with the $B\to D^{*}_0$ transition accompanied by the emission of a light pseudoscalar meson ($\pi, K$), the other is associated with the $B\to \pi (K)$ transition accompanied by the scalar meson $D^*_0$ emission. We can find that the former gives the dominant contribution. The other two decay channels $B^{0}\to D^{\ast-}_{0}\pi^{+}$ and $B^{0}\to D^{\ast-}_{0}K^{+}$  only receive one kind  of Feynman diagram contribution to the their branching ratios, which is associated with the $B\to D^{*}_0$ transition accompanied by a light pseudoscalar meson ($\pi, K$) emission.  In any case, the contributions from the vertex corrections and the hard spectator-scatterings are destructive with each other. In view of the large uncertainty from the decay constant $f_{D^{\ast}_{0}}$ as mentioned before, two decay constant values are used in our calculations, which are shown in Table \ref{D2300pik}. For each decay, the upper line is the result corresponding to $f_{D^{\ast}_{0}}=0.078$ GeV, and the lower is the result corresponding to $f_{D^{\ast}_{0}}=0.103$ GeV. We can find that the branching ratios are sensitive to the decay constant $f_{D^{\ast}_{0}}$. Our predictions are consistent with other theoretical calculations, such as the PQCD approach \cite{Wang:2018fai}, the ISGW quark model \cite{Cheng:2003id,Katoch:1995hr}. Certainly, the branching ratio of the decay $B^{+}\to \bar{D}^{\ast0}_{0}\pi^{+}$ was also calculated using the CLFQM in previous \cite{Cheng:2003sm}, which is agreement with our result.

\begin{table}[H]
\caption{The CLFQM predictions for the branching ratios of the decays $B\to D^*_0 \pi (K)\to D\pi \pi (K)$, where the upper (lower) line is corresponding to $f_{D^{\ast}_{0}}=0.078 (0.103)$ GeV for each decay. The first and second uncertainties are from the decay constants of $B$ and $D^*_0$.}
\begin{center}
\scalebox{0.8}{
\begin{tabular}{|c |c |c|c |}
\hline\hline
&This work&PQCD \cite{Wang:2018fai}&Data\\
\hline\hline
&$2.79^{+0.85+1.35}_{-0.68-1.18}$&$$&$6.1\pm0.6\pm0.9\pm1.6$ Belle \cite{belle}\\
$10^{-4}\times\mathcal{B} r(B^{+}\to \bar{D}^{\ast0}_{0}\pi^{+}\to D^{-}\pi^{+}\pi^{+})$&$$&$5.95^{+3.14}_{-2.32}$&$6.8\pm0.3\pm0.4\pm2.0$ BaBar \cite{babar2}\\
&$5.06^{+0.99+1.99}_{-0.88-1.92}$&$$& $5.78\pm0.08\pm0.06\pm0.09$ LHCb \cite{LHCb:2016lxy} \\
\hline
&$2.50^{+0.77+1.23}_{-0.82-1.07}$&$$&$0.60\pm0.13\pm0.15\pm0.22 $ Belle \cite{Belle:2006wbx} \\
$10^{-4}\times\mathcal{B} r({B}^{0}\to D^{\ast-}_{0}\pi^{+}\to \bar{D^{0}}\pi^{-}\pi^{+})$&$$&$2.85^{+1.65}_{-1.18}$&$0.77\pm0.05\pm0.03\pm0.03$ LHCb \cite{LHCb:2015klp}\\
&$4.54^{+0.89+1.81}_{-0.80-1.74}$&$$& $0.80\pm0.05\pm0.08\pm0.04 $ LHCb \cite{LHCb:2015klp}\\
\hline
$10^{-5}\times\mathcal{B} r(B^{+}\to \bar{D}^{\ast0}_{0} K^{+}\to D^{-}\pi^{+}K^{+})$ &$2.13^{+0.65+1.05}_{-0.52-0.91}$&$4.65^{+2.46}_{-1.85}$&$0.61\pm0.19\pm0.05\pm0.14\pm0.04 $ LHCb \cite{LHCb:2015eqv}\\
&$4.05^{+0.78+1.57}_{-0.70-1.51}$&$$&$$\\
\hline
$10^{-5}\times\mathcal{B} r({B}^{0}\to D^{\ast-}_{0}K^{+}\to \bar{D^{0}}\pi^{-}K^{+})$ &$1.96^{+0.61+0.97}_{-0.49-0.84}$&$2.38^{+1.31}_{-0.98}$&$1.77\pm0.26\pm0.19\pm0.67\pm0.20 $ LHCb \cite{LHCb:2015tsv}\\
&$3.57^{+0.70+1.42}_{-0.63-1.37}$&$$&$$\\
\hline\hline
\end{tabular}\label{D2300pik2}}
\end{center}
\end{table}

The branching ratio of the quasi-two-body decay $B\to D^*_0P\to D\pi P$ can be obtained from the corresponding two body decay $B\to D^*_0P$ under the narrow width approximation $\mathcal{B} r(B\to D^{\ast}_{0}P\to D\pi P)=\mathcal{B} r(B\to D^{\ast}_{0}P)\mathcal{B} r(D^{\ast}_{0}\to D\pi)$,
where $P$ refers to a light pseudoscalar meson ($\pi, K$).
Assuming the $D^{\ast}_{0}$ state decays essentially into D$\pi$, we have $\mathcal{B} r(\bar{D}^{\ast0}_{0}\to D^{-}\pi^{+})=\mathcal{B} r(D^{\ast-}_{0}\to \bar{D}^{0}\pi^{-})$ $=\frac{2}{3}$ from the Clebsch-Gordan coefficients. The branching ratios of these quasi-two-body decays are collected in Table \ref{D2300pik2}, where the PQCD results and the data are also listed for comparison. One can find that if taken the bigger decay constant $f_{D^*_0}=0.103$ GeV, the branching fraction for the decay  $B^{+}\to \bar{D}^{\ast0}_{0}\pi^{+}\to D^{-}\pi^{+}\pi^{+}$ can agree with the data from Belle \cite{belle}, BaBar \cite{babar2} and LHCb \cite{LHCb:2016lxy} within errors. While for the decay ${B}^{0}\to D^{\ast-}_{0}K^{+}\to \bar{D}^{0}\pi^{-}K^{+}$, if taken the smaller decay constant $f_{D^*_0}=0.078$ GeV, our prediction can explain the LHCb measurement \cite{LHCb:2015tsv}. For the other two quasi-two-body decays, the predictions for their branching ratios are much larger than the present data. Similar situation exists for the comparison between the PQCD results and the measurements.  Another divergence is that our results for the charged (neutral) channels  by using the bigger (smaller) decay constant $f_{D^*_0}=0.103\ (0.078)$ GeV can be consistent well with the PQCD results.  Further experimental and theoretical researches are needed to clarify these divergences and puzzles.

\begin{table}[H]
\caption{The CLFQM predictions for the branching ratios ($10^{-3}$) of the decays $B \to D^{\ast}_{0}D_{(s)}, D^{\ast}_{0}D^*_{(s)}, D^{\ast}_{0}\rho, D^{\ast}_{0} K^*$. The labels LO, VC, HSSC, NLO and the error sources are the same with those given in Table \ref{D2300pik}. Other theoretical predictions are also listed for comparison.}
\begin{center}
\scalebox{0.75}{
\begin{tabular}{|c|c|c|c|c|c|c|}
\hline\hline
  &LO & VC&HSSC&NLO&ISGW2 \cite{Cheng:2003id}&ISGW \cite{Katoch:1995hr}\\
\hline\hline
$\mathcal{B} r(B^{+}\to \bar{D}^{\ast0}_{0}D_{s}^{+})$&$1.54^{+0.47+0.76}_{-0.38-0.66}$&$1.52^{+0.47+0.75}_{-0.38-0.65}$&$1.52^{+0.47+0.75}_{-0.38-0.65}$&$1.51^{+0.46+0.74}_{-0.37-0.65}$&$0.80$&$2.7$\\
&$2.79^{+0.55+1.11}_{-0.49-1.07}$&$2.76^{+0.54+1.10}_{-0.49-1.06}$&$2.76^{+0.55+1.10}_{-0.49-1.06}$&$2.74^{+0.54+1.09}_{-0.48-1.05}$&$$&$$\\
$\mathcal{B} r({B}^{0}\to D^{\ast-}_{0}D_{s}^{+})$&$1.42^{+0.44+0.70}_{-0.35-0.61}$&$1.41^{+0.43+0.70}_{-0.35-0.61}$&$1.41^{+0.44+0.70}_{-0.35-0.61}$&$1.40^{+0.43+0.69}_{-0.34-0.60}$&$0.73$&$2.6$\\
&$2.59^{+0.51+1.03}_{-0.46-0.99}$&$2.56^{+0.51+1.02}_{-0.45-0.98}$&$2.53^{+0.50+1.01}_{-0.45-0.97}$&$2.51^{+0.49+1.00}_{-0.44-0.96}$&$$&$$\\
\hline
$\mathcal{B} r(B^{+}\to \bar{D}^{\ast0}_{0}D_{s}^{\ast+})$&$1.71^{+0.53+0.85}_{-0.42-0.74}$&$1.70^{+0.52+0.84}_{-0.42-0.73}$&$1.73^{+0.53+0.85}_{-0.43-0.74}$&$1.71^{+0.53+0.84}_{-0.42-0.73}$&$0.35$&$-$\\
&$3.11^{+0.61+1.24}_{-0.55-1.19}$&$3.08^{+0.61+1.23}_{-0.54-1.18}$&$3.14^{+0.62+1.25}_{-0.55-1.20}$&$3.11^{+0.61+1.24}_{-0.55-1.19}$&$$&$$\\
$\mathcal{B} r({B}^{0}\to D^{\ast-}_{0}D_{s}^{\ast+})$&$1.59^{+0.49+0.78}_{-0.39-0.68}$&$1.57^{+0.49+0.78}_{-0.39-0.68}$&$1.60^{+0.49+0.80}_{-0.39-0.69}$&$1.58^{+0.49+0.78}_{-0.39-0.68}$&$0.32$&$-$\\
&$2.89^{+0.57+1.15}_{-0.51-1.11}$&$2.86^{+0.56+1.14}_{-0.50-1.10}$&$2.86^{+0.56+1.14}_{-0.50-1.10}$&$2.83^{+0.56+1.13}_{-0.50-1.09}$&$$&$$\\
\hline
$\mathcal{B} r(B^{+}\to \bar{D}^{\ast0}_{0}\rho^{+})$&$0.81^{+0.28+0.44}_{-0.22-0.38}$&$0.85^{+0.29+0.47}_{-0.22-0.40}$&$0.89^{+0.29+0.46}_{-0.23-0.40}$&$0.94^{+0.30+0.48}_{-0.24-0.42}$&$1.30$&$-$\\
&$1.47^{+0.32+0.65}_{-0.29-0.61}$&$2.03^{+0.38+0.77}_{-0.34-0.75}$&$1.63^{+0.34+0.68}_{-0.30-0.65}$&$2.21^{+0.39+0.80}_{-0.36-0.78}$&$$&$$\\
$\mathcal{B} r({B}^{0}\to D^{\ast-}_{0}\rho^{+})$&$0.94^{+0.29+0.46}_{-0.23-0.40}$&$0.97^{+0.30+0.48}_{-0.24-0.42}$&$0.95^{+0.29+0.47}_{-0.23-0.41}$&$0.99^{+0.30+0.49}_{-0.24-0.42}$&$0.64$&$-$\\
&$1.70^{+0.34+0.68}_{-0.30-0.65}$&$1.77^{+0.35+0.71}_{-0.31-0.68}$&$1.66^{+0.33+0.66}_{-0.29-0.64}$&$1.73^{+0.34+0.69}_{-0.31-0.66}$&$$&$$\\
\hline
$\mathcal{B} r(B^{+}\to \bar{D}^{\ast0}_{0}K^{\ast+})$&$0.045^{+0.016+0.025}_{-0.012-0.021}$&$0.047^{+0.016+0.026}_{-0.013-0.022}$&$0.049^{+0.016+0.026}_{-0.013-0.022}$&$0.052^{+0.017+0.027}_{-0.013-0.023}$&$-$&$-$\\
&$0.082^{+0.018+0.037}_{-0.016-0.034}$&$0.116^{+0.022+0.044}_{-0.020-0.043}$&$0.090^{+0.019+0.038}_{-0.017-0.036}$&$0.126^{+0.022+0.045}_{+0.020+0.044}$&$$&$$\\
$\mathcal{B} r({B}^{0}\to D^{\ast-}_{0}K^{\ast+})$&$0.054^{+0.017+0.026}_{-0.013-0.023}$&$0.056^{+0.017+0.028}_{-0.014-0.024}$&$0.054^{+0.017+0.027}_{-0.013-0.023}$&$0.056^{+0.017+0.028}_{-0.014-0.024}$&$-$&$-$\\
&$0.097^{+0.019+0.039}_{-0.017-0.037}$&$0.101^{+0.020+0.040}_{-0.018-0.039}$&$0.095^{+0.019+0.038}_{-0.017-0.036}$&$0.099^{+0.020+0.040}_{-0.018-0.038}$&$$&$$\\
\hline
$\mathcal{B} r(B^{+}\to \bar{D}^{\ast0}_{0}D^{+})$&$0.052^{+0.016+0.026}_{-0.013-0.022}$&$0.051^{+0.016+0.025}_{-0.013-0.022}$&$0.052^{+0.016+0.026}_{-0.013-00.22}$&$0.051^{+0.016+0.025}_{-0.013-0.022}$&$-$&$0.114$\\
&$0.094^{+0.019+0.038}_{-0.017-0.036}$&$0.093^{+0.018+0.037}_{-0.016-0.036}$&$0.094^{+0.019+0.037}_{-0.017-0.036}$&$0.092^{+0.018+0.037}_{-0.016-0.035}$&$$&$$\\
$\mathcal{B} r({B}^{0}\to D^{\ast-}_{0}D^{+})$&$0.048^{+0.015+0.024}_{-0.012-0.021}$&$0.047^{+0.015+0.023}_{-0.012-0.020}$&$0.048^{+0.015+0.024}_{-0.012-0.021}$&$0.047^{+0.015+0.023}_{-0.012-0.020}$&$-$&$0.111$\\
&$0.087^{+0.017+0.035}_{-0.015-0.033}$&$0.086^{+0.017+0.034}_{-0.015-0.033}$&$0.086^{+0.017+0.034}_{-0.015-0.033}$&$0.085^{+0.017+0.034}_{-0.015-0.032}$&$$&$$\\
\hline
$\mathcal{B} r(B^{+}\to \bar{D}^{\ast0}_{0}D^{\ast+})$&$0.074^{+0.023+0.037}_{-0.018-0.032}$&$0.073^{+0.023+0.036}_{-0.018-0.031}$&$0.075^{+0.023+0.037}_{-0.019-0.032}$&$0.074^{+0.023+0.037}_{-0.018-0.032}$&$-$&$-$\\
&$0.135^{+0.027+0.054}_{-0.024-0.052}$&$0.133^{+0.026+0.053}_{-0.023-0.051}$&$0.137^{+0.027+0.054}_{-0.024-0.052}$&$1.340^{+0.027+0.054}_{-0.024-0.052}$&$$&$$\\
$\mathcal{B} r({B}^{0}\to D^{\ast-}_{0}D^{\ast+})$&$0.069^{+0.021+0.034}_{-0.017-0.030}$&$0.068^{+0.021+0.033}_{-0.017-0.029}$&$0.070^{+0.022+0.034}_{-0.017-0.030}$&$0.069^{+0.021+0.034}_{-0.017-0.029}$&$-$&$-$\\
&$0.125^{+0.024+0.050}_{-0.022-0.048}$&$0.123^{+0.024+0.049}_{-0.022-0.047}$&$0.124^{+0.025+0.050}_{-0.022-0.048}$&$0.122^{+0.024+0.049}_{-0.022-0.047}$&$$&$$\\
\hline\hline
\end{tabular}\label{D23003}}
\end{center}
\end{table}

If replaced the light pseudoscalar mesons $\pi,K$ in the final states with the vector meson $\rho, K^*$ and the charmed meson $D^{(*)},D^{(*)}_s$, the branching ratios for the conrresponding decay channels are listed in Table \ref{D23003}. One can find that our predictions by using the bigger decay constant $f_{D^*_0}=0.103$ GeV are consistent with those given in the ISGW model, it is because that the form factor of the transition $B\to D^*_0$ at maximum momentum transfer obtained in the CLFQM with $f_{D^*_0}=0.103$ GeV is about 0.30, which is almost equal to the value calculated in the ISGW model \cite{Katoch:1995hr}, while much larger than
0.18 given in the ISGW2 model \cite{Cheng:2003id}. The differences between the ISGW and ISGW2 models are mainly from the $q^2$-dependence of the form factor.
About twteen years ago, Chua \cite{Chua:2003ac} studied the decay $B^{+}\to \bar{D}^{\ast0}_{0}\rho^{+}$ in the CLFQM approach and obtained its branching fraction about $1.7\times10^{-3}$, which is consistent with our result by taking $f_{D^*_0}=0.103$ GeV.

\begin{table}[H]
	\caption{The CLFQM predictions for the branching ratios ($10^{-4}$) of  the decays  $B_{s}\to  D^{\ast}_{s0}\pi(K), D^{\ast}_{s0}\rho(K^*)$. The labels LO, VC, HSSC, NLO and the error sources are the same with those given in Table \ref{D2300pik}. Other theoretical predictions are also listed for comparison.}
	\begin{center}
		\scalebox{0.90}{
			\begin{tabular}{|c|c|c|c|c|}
				\hline\hline
				& $\mathcal{B} r( B_s\to D^{\ast-}_{s0}\pi^{+})$ &$\mathcal{B} r( B_s\to D^{\ast-}_{s0}K^+)$&$\mathcal{B} r( B_s\to D^{\ast-}_{s0}\rho^+)$&$\mathcal{B} r( B_s\to D^{\ast-}_{s0}K^{*+})$\\
				\hline\hline
				LO&$5.06^{+0.83+2.09}_{-0.75-1.83}$&$0.40^{+0.07+0.17}_{-0.06-0.15}$&$12.87^{+2.11+5.31}_{-1.91-4.66}$&$0.74^{+0.12+0.30}_{-0.11-0.27}$\\
				VC&$5.27^{+0.86+2.17}_{-0.78-1.91}$&$0.42^{+0.07+0.17}_{-0.06-0.15}$&$13.40^{+2.20+5.53}_{-1.99-4.86}$&$0.77^{+0.13+0.32}_{-0.11-0.28}$\\
				HSSC&$4.96^{+0.81+2.05}_{-0.74-1.80}$&$0.38^{+0.06+0.16}_{-0.06-0.14}$&$12.59^{+2.06+5.20}_{-1.87-4.56}$&$0.72^{+0.12+0.30}_{-0.11-0.26}$\\
				NLO&$5.17^{+0.85+2.13}_{-0.77-1.87}$&$0.40^{+0.07+0.16}_{-0.06-0.14}$&$13.12^{+2.15+5.42}_{-1.95-4.76}$&$0.75^{+0.12+0.31}_{-0.11-0.28}$\\
				PQCD \cite{Zhang:2019pax}&$5.49^{+2.64+0.41+0.35}_{-1.68-0.27-0.35}$&$0.51^{+0.06+0.01+0.01}_{-0.04-0.01-0.01}$&$17.7^{+8.5+1.3+1.2}_{-5.3-0.8-1.1}$&$1.01^{+0.44+0.06+0.05}_{-0.31-0.06-0.07}$\\
				LSCR \cite{Li:2009wq}&$5.2^{+2.5}_{-2.1}$&$0.4^{+0.2}_{-0.2}$&$13^{+6}_{-5}$&$0.8^{+0.4}_{-0.3}$\\
				RQM \cite{Kramer2}&$9$&$0.7$&$22$&$1.2$\\
				NRQM \cite{Albertus:2014bfa}&$10$&$0.9$&$27$&$16$\\
				ISGW2 \cite{Cheng:2003id}&$3.3$&$-$&$8.3$&$-$\\
				\hline\hline
			\end{tabular}\label{D2317pik}}
	\end{center}
\end{table}

Taking $D^*_{s0}$ as a $c\bar s$ meson, we calclate the branching ratios of the decays $B_{(s)}\to D^*_{s0}M$ with $M$ being a pseudoscalar meson ($\pi, K, D, D_s$) or a vector meson $(\rho, K^*, D^*, D^*_s)$ in the CLFQM, which are listed in Tables \ref{D2317pik}, \ref{D2317D} and \ref{D2317D2}. From Table \ref{D2317pik},  one can find that our predictions for the decays $B_{s}\to  D^{\ast}_{s0}\pi(K), D^{\ast}_{s0}\rho(K^*)$ are consistent well with those calculated in the LCSR \cite{Li:2009wq} and  the PQCD approach \cite{Zhang:2019pax} within errors, while (much) smaller than those given by the RQM \cite{Kramer2} and the nonrelativistic quark model (NRQM) \cite{Albertus:2014bfa}. Especially, for the pure annihilation decay $ B_s\to D^{*-}_{s0}K^{*+}$, its branching fraction reaches up to $10^{-3}$ predicted by the NRQM \cite{Albertus:2014bfa}, which seems too large compared to other theoretical results. These divergences can be clarified by the future LHCb and Super KEKB experiments. The branching ratios of the decays $B_s\to D^{*-}_{s0}\pi^+(\rho^+)$ were also calculated in the ISGW2 model \cite{Cheng:2003id}, which are smaller than our results. It is because of the difference from the form factor of the transition $B_s\to D^{*}_{s0}$ and its $q^2$-dependence. It is similar for the decays $B^0\to D^{*-}_0\pi^+(\rho^+)$.

\begin{table}[H]
\caption{The CLFQM predictions for the branching ratios ($10^{-4}$) of the decays $B^+\to D^{\ast+}_{s0}\bar D^{(*)0}$ and $B^0\to D^{\ast+}_{s0}D^{(*)-}$. The labels LO, VC, HSSC, NLO and the error sources are the same with those given in Table \ref{D2300pik}. Other theoretical predictions and data are also listed for comparison.  }
\begin{center}
\scalebox{0.90}{
\begin{tabular}{|c|c|c|c|c|}
\hline\hline
&$\mathcal{B} r(B^+\to D^{\ast+}_{s0}\bar D^0)$&$\mathcal{B} r(B^+\to D^{\ast+}_{s0}\bar D^{*0})$&$\mathcal{B} r(B^0\to D^{\ast+}_{s0}D^-)$&$\mathcal{B} r(B^0\to D^{\ast+}_{s0}D^{*-})$\\
\hline\hline
LO&$13.81^{+0.04+0.19}_{-0.21-0.26}$&$13.99^{+0.12+0.87}_{-0.30-0.95}$&$12.80^{+0.04+0.17}_{-0.19-0.24}$&$12.97^{+0.11+0.80}_{-0.28-0.88}$\\
VC&$13.66^{+0.04+0.18}_{-0.21-0.26}$&$13.83^{+0.12+0.86}_{-0.30-0.94}$&$12.66^{+0.04+0.17}_{-0.19-0.24}$&$12.83^{+0.11+0.80}_{-0.28-0.87}$\\
HSSC&$13.62^{+0.04+0.18}_{-0.21-0.25}$&$13.95^{+0.12+0.87}_{-0.30-0.95}$&$12.62^{+0.04+0.17}_{-0.19-0.24}$&$12.96^{+0.11+0.80}_{-0.28-0.88}$\\
NLO&$13.47^{+0.04+0.18}_{-0.20-0.25}$&$13.80^{+0.12+0.86}_{-0.30-0.94}$&$12.48^{+0.04+0.17}_{-0.19-0.23}$&$12.81^{+0.11+0.80}_{-0.28-0.87}$\\
PQCD \cite{Zhang:2021bcr}&$11.2^{+4.0+0.3+0.4}_{-2.8-0.2-0.4}$&$18.3^{+7.1+2.7+0.7}_{-5.4-1.7-0.5}$&$10.5^{+4.5+0.4+0.4}_{-3.0-0.2-0.4}$&$15.9^{+7.0+2.4+0.6}_{-4.9-1.4-0.5}$\\
FH \cite{Faessler:2007cu}&$10.3\pm1.4$&$5.0\pm0.7$&$9.6\pm1.3$&$4.7\pm0.6$\\
TM \cite{Liu:2022dmm}&$6.77\pm1.9$&$12.10\pm3.39$&$6.37\pm1.78$&$8.89\pm2.49$\\
Data \cite{pdg22}&$8.0^{+1.6}_{-1.3}$&$9.0^{+7.0}_{-7.0}$&$10.6^{+1.6}_{-1.6}$&$15.0^{+6.0}_{-6.0}$\\
\hline\hline
\end{tabular}\label{D2317D}}
\end{center}
\end{table}

In Table \ref{D2317D}, all the predictions from the different theories, including the PQCD approach \cite{Zhang:2021bcr}, the factorization hypothesis (FH) \cite{Faessler:2007cu} and the triangle mechanism (TM) \cite{Liu:2022dmm}, show that the branching ratios of the charged decay $B^+\to D^{\ast+}_{s0}\bar D^{(*)0}$ are slightly larger than those of the corresponding neutral decays $B^0\to D^{\ast+}_{s0}D^{(*)-}$. It is just contrary to the case of the data \cite{pdg22}. Certainly, there still exist large errors in the experimental results, especially for the branching ratios of the decays with the vector meson $D^{*}$ involved. We expect more accurate measurements in the future LHCb and Super KEKB experiments. Theoretically, the decays $B^+\to D^{\ast+}_{s0}\bar D^{(*)0}$ and $B^0\to D^{\ast+}_{s0}D^{(*)-}$ have the same CKM matrix elements and Wilson coefficients for the factorizable emission amplitudes, which provide the dominant contributions to their branching ratios. Furthermore, there exist similar transition form factors for isospin symmetry among these channels. So these four decays should have similar branching ratios. From Table \ref{D2317D2},  one can find that the branching ratios of the decays $B_s\to D^{*}_{s0}D^{(*)}_s, D^{*}_{s0}D^{(*)}$ are consistent with those given in the PQCD approach \cite{Zhang:2019pax} and the RQM \cite{Kramer2}, while much smaller than those obtained within the LCSR \cite{Li:2009wq}. Further experimental and theoretical researches are needed to clarify these divergences. For the decays $B_s\to D^{\ast-}_{s0}D^{(*)+}$, their branching ratios are much smaller than those of other four channels mainly because of the smaller CKM matrix element $V_{cd}$ compared with $V_{cs}$, that is to say there exists a suppressed factor $|V_{cd}/V_{cs}|^2 \approx 0.05$ for the decays $B_s\to D^{\ast-}_{s0}D^{(*)+}$ compared to other four channels.

\begin{table}[H]
	\caption{The branching ratios ($10^{-3}$) of the  decays $B_{s}\to D^{\ast}_{s0}D^{(*)}_s,D^{\ast}_{s0}D^{(*)}$. The labels LO, VC, HSSC, NLO and the error sources are the same with those given in Table \ref{D2300pik}. Other theoretical predictions are also listed for comparison. }
	\begin{center}
		\scalebox{0.7}{
			\begin{tabular}{|c|c|c|c|c|c|c|c|}
				\hline\hline
				&LO & VC&HSSC&NLO &PQCD \cite{Zhang:2021bcr}&RQM \cite{Kramer2} &LSCR \cite{Li:2009wq}\\
				\hline\hline
				$\mathcal{B} r(B_s\to D^{\ast-}_{s0}D^+_s)$&$1.96^{+0.32+0.81}_{-0.29-0.71}$&$1.94^{+0.32+0.80}_{-0.29-0.70}$&$1.94^{+0.32+0.80}_{-0.29-0.70}$&$1.92^{+0.31+0.79}_{-0.28-0.69}$&$2.1^{+0.9+0.3+0.1}_{-0.6-0.1-0.1}$&$1.1$&$13^{+7}_{-5}$\\
				
				$\mathcal{B} r(B_s\to D^{\ast-}_{s0}D^{*+}_s)$&$2.18^{+0.36+0.90}_{-0.32-0.79}$&$2.16^{+0.35+0.89}_{-0.32-0.78}$&$2.16^{+0.35+0.89}_{-0.32-0.78}$&$2.14^{+0.35+0.88}_{-0.32-0.78}$&$1.8^{+0.9+0.1+0.1}_{-0.6-0.1-0.1}$&$2.3$&$6.0^{+2.9}_{-2.4}$\\
				
				$\mathcal{B} r(B_s\to D^{\ast-}_{s0}D^+)$&$0.066^{+0.011+0.027}_{-0.010-0.024}$&$0.065^{+0.010+0.027}_{-0.011-0.024}$&$0.065^{+0.010+0.027}_{-0.011-0.024}$&$0.064^{+0.011+0.027}_{-0.010-0.023}$&$0.065^{+0.034+0.006+0.002}_{-0.021-0.004-0.002}$&$-$&$0.5^{+0.2}_{-0.2}$\\
				
				$\mathcal{B} r(B_s\to D^{\ast-}_{s0}D^{*+})$&$0.095^{+0.016+0.039}_{-0.014-0.034}$&$0.093^{+0.015+0.038}_{-0.014-0.034}$&$0.094^{+0.015+0.039}_{-0.014-0.034}$&$0.092^{+0.015+0.038}_{-0.014-0.033}$&$0.050^{+0.026+0.004+0.002}_{-0.017-0.002-0.001}$&$-$&$0.2^{+0.1}_{-0.1}$\\
				
				$\mathcal{B} r(B_s\to D^{\ast+}_{s0}D^-_s)$&$1.29^{+0.00+0.06}_{-0.01-0.06}$&$1.28^{+0.00+0.06}_{-0.01-0.06}$&$1.24^{+0.00+0.06}_{-0.01-0.06}$&$1.22^{+0.00+0.06}_{-0.01-0.06}$&$1.11^{+0.56+0.02+0.04}_{-0.37-0.04-0.04}$&$-$&$-$\\
				
				$\mathcal{B} r(B_s\to D^{\ast+}_{s0}D^{*-}_s)$&$1.28^{+0.01+0.06}_{-0.02-0.11}$&$1.27^{+0.01+0.06}_{-0.01-0.06}$&$1.26^{+0.01+0.06}_{-0.02-0.11}$&$1.24^{+0.01+0.06}_{-0.02-0.11}$&$1.48^{+0.69+0.05+0.06}_{-0.46-0.07-0.05}$&$-$&$-$\\
				\hline\hline
			\end{tabular}\label{D2317D2}}
	\end{center}
\end{table}

 In the quark model the axial-vector mesons exist in two types of spectroscopic states, $^3P_1(J^{PC}=1^{++})$ and $^1P_1(J^{PC}=1^{+-})$. In some cases the physical particles are the mixture of these two types of states. For example, $K_1(1270)$ and $K_1(1400)$ are considered as the mixture of $K_{1A}$ and $K_{1B}$ for the mass difference of the strange and light quarks. Similarly, the charm-strange mesons $D_{s1}$ and $D_{s1}^\prime$ are usually written as the mixture of the states $^1D_{s1}$ and $^3D_{s1}$, which are given in Eq. (\ref{mixing3}). The quark potential model determined the mixing angle  $\theta_s\approx7^\circ$ \cite{Cheng:2003id}. So we use $\theta_s=7^\circ$ to calculate the branching ratios of the decays $B_s\to D_{s1}^{(\prime)}P(V)$ with $P$ and $V$ being the pseudoscalar mesons ($\pi, K, D, D_s$) and the vector mesons $(\rho, K^*, D^*, D^*_s)$, respectivley, which are listed in Table \ref{mix state} with the results given in the RQM, the NRQM and the ISGW2 models \cite{Kramer2,Albertus:2014bfa,Cheng:2003id} for comparison. The following points can be found
 \begin{itemize}
 	\item  It is interesting that our predictions for the decays $B_s\to D_{s1}P(V)$ are consistent well with the results obtained in the RQM \cite{Kramer2} and NRQM \cite{Albertus:2014bfa} models, while those for most of the channels $B_s\to D_{s1}^\prime P(V)$ are about $3$ times smaller than the RQM calculations. Certainly, the decays $B_s\to D_{s1}\pi(\rho)$ and $B_s\to D_{s1}^\prime\pi(\rho)$ have been researched by using the ISGW2 model \cite{Cheng:2003id} about twenty years ago. We argue that the mixing formula between $|^1D_{s1}\rangle$ and $|^3D_{s1}\rangle$ used in Ref. \cite{Cheng:2003id} is incorrect, which induced $\mathcal{B}r(B_s\to D_{s1}^\prime\pi(\rho))$ are larger than $\mathcal{B}r(B_s\to D_{s1}\pi(\rho))$. It is just contrary with other theoretical predictions. That is to say the values of the branching raitos $\mathcal{B}r(B_s\to D_{s1}^\prime\pi(\rho))$ and $\mathcal{B}r(B_s\to D_{s1}\pi(\rho))$ should be exchanged with each other under the correct mixing formula shown in Eq. (\ref{mixing3}). Since many of these decays have large branching ratios, which lie in the range $\mathcal{O}(10^{-5})\sim \mathcal{O}(10^{-3})$, we expect that the LHCb and Super KEKB experiments can clarify the differences between these results in the future.
 	\item  The branching ratios of the CKM-favored decays $B_s\to D^{(')}_{s1} D^{(*)}_s$  and $B_s\to D^{(')}_{s1} \pi(\rho)$, which are associated with the CKM matrix elements $V_{cs}$ and $V_{ud} (\sim1)$, respectively, are much larger than those of the CKM-suppressed channels $B_s\to D^{(')}_{s1} D^{(*)}$ and $B_s\to D^{(')}_{s1} K^{(*)}$, which are associated with the CKM matrix elements $V_{cd}$ and $V_{us} (\approx0.22)$, respectively. It shows a clear hierarchical relationship for the
 	branching raitos of these color-favored decay modes,
 	\begin{footnotesize}
 		\begin{eqnarray}
 			\mathcal{B}r(B_s\to D^{(')}_{s1} D^{(*)}_s) \gg \mathcal{B}r(B_s\to D^{(')}_{s1} D^{(*)}),\;\; \mathcal{B}r(B_s\to D^{(')}_{s1} \pi(\rho)) \gg \mathcal{B}r(B_s\to D^{(')}_{s1} K^{(*)}).
 		\end{eqnarray}
 	\end{footnotesize}
\begin{table}[H]
	\caption{The branching ratios ($10^{-3}$) of the decays $B_{s}\to D^{(\prime)}_{s1}P(V)$. The labels LO, VC, HSSC, NLO and the error sources are the same with those given in Table \ref{D2300pik}. The results given in the RQM, the NRQM and the ISGW2 models are also listed for comparison.  }
	\begin{center}
		\scalebox{0.60}{
			\begin{tabular}{|c|c|c|c|c|c|c|c|}
				\hline\hline
				&LO& VC&HSSC&NLO &RQM \cite{Kramer2}&NRQM \cite{Albertus:2014bfa}&ISGW2 \cite {Cheng:2003id}\\
				\hline\hline
				$\mathcal{B} r( B_s\to D_{s1}^-D_{s}^{+})$&$6.55^{+0.56+0.12+1.36}_{-0.55-0.18-1.29}$&$6.48^{+0.56+0.12+1.34}_{-0.55-0.17-1.28}$&$6.44^{+0.56+0.12+1.36}_{-0.55-0.17-1.28}$&$6.38^{+0.56+0.12+1.34}_{-0.54-0.17-1.27}$&$3.0$&$-$&$-$\\
				$\mathcal{B} r( B_s\to D_{s1}^-D^{\ast+}_{s})$&$8.45^{+0.72+0.24+1.59}_{-0.70-0.32-1.53}$&$8.48^{+0.72+0.25+1.58}_{-0.70-0.33-1.52}$&$8.29^{+0.71+0.23+1.59}_{-0.69-0.31-1.52}$&$8.32^{+0.71+0.24+1.58}_{-0.70-0.32-1.52}$&$5.9$&$-$&$-$\\
				$\mathcal{B} r(B_s\to D_{s1}^-\pi^{+})$&$1.67^{+0.14+0.03+0.35}_{-0.14-0.04-0.33}$&$1.74^{+0.15+0.03+0.36}_{-0.15-0.05-0.34}$&$1.64^{+0.14+0.03+0.34}_{-0.14-0.04-0.33}$&$1.70^{+0.15+0.03+0.36}_{-0.15-0.01-0.34}$&$1.9$&$1.5$&$0.52$\\
				$\mathcal{B} r( B_s\to D_{s1}^-\rho^{+})$&$4.49^{+0.39+0.10+0.90}_{-0.38-0.14-0.86}$&$4.68^{+0.40+0.11+0.94}_{-0.39-0.15-0.90}$&$4.40^{+0.38+0.10+0.90}_{-0.38-0.13-0.86}$&$4.59^{+0.40+0.10+0.94}_{+0.39+0.14+0.89}$&$4.9$&$3.6$&$1.3$\\				
				$\mathcal{B} r(B_s\to D_{s1}^-K^{+})$&$0.13^{+0.01+0.00+0.03}_{-0.01-0.00-0.03}$&$0.14^{+0.01+0.00+0.03}_{-0.01-0.00-0.03}$&$0.13^{+0.01+0.00+0.03}_{-0.01-0.00-0.03}$&$0.14^{+0.01+0.00+0.03}_{-0.01-0.00-0.03}$&$0.14$&$0.12$&$-$\\
				$\mathcal{B} r( B_s\to D_{s1}^-K^{\ast+})$&$0.26^{+0.02+0.01+0.05}_{-0.02-0.01-0.05}$&$0.27^{+0.02+0.01+0.05}_{-0.02-0.01-0.05}$&$0.26^{+0.02+0.01+0.05}_{-0.02-0.01-0.05}$&$0.27^{+0.02+0.01+0.05}_{-0.02-0.01-0.05}$&$0.26$&$0.20$&$-$\\
				$\mathcal{B} r( B_s\to D_{s1}^-D^{+})$&$0.22^{+0.02+0.00+0.05}_{-0.02-0.01-0.04}$&$0.22^{+0.02+0.00+0.05}_{-0.02-0.01-0.04}$&$0.22^{+0.02+0.00+0.05}_{-0.02-0.01-0.04}$&$0.21^{+0.02+0.00+0.05}_{-0.02-0.01-0.04}$&$-$&$-$&$-$\\
				$\mathcal{B} r(B_s\to D_{s1}^-D^{\ast+})$&$0.37^{+0.03+0.01+0.07}_{-0.03-0.01-0.07}$&$0.37^{+0.03+0.01+0.07}_{-0.03-0.01-0.07}$&$0.36^{+0.03+0.01+0.07}_{-0.03-0.01-0.07}$&$0.36^{+0.03+0.01+0.07}_{-0.03-0.01-0.07}$&$-$&$-$&$-$\\
				\hline
				$\mathcal{B} r( B_s\to D_{s1}^{'-}D_{s}^{+})$&$0.27^{+0.10+0.04+0.29}_{-0.08-0.03-0.19}$&$0.26^{+0.10+0.04+0.29}_{-0.08-0.03-0.19}$&$0.32^{+0.11+0.04+0.31}_{-0.09-0.03-0.21}$&$0.31^{+0.11+0.04+0.31}_{-0.09-0.03-0.21}$&$0.54$&$-$&$-$\\
				$\mathcal{B} r(B_s\to D_{s1}^{'-}D^{\ast+}_{s})$&$0.42^{+0.15+0.01+0.30}_{-0.12-0.00-0.18}$&$0.43^{+0.16+0.02+0.30}_{-0.13-0.00-0.18}$&$0.46^{+0.16+0.01+0.33}_{-0.13-0.01-0.21}$&$0.46^{+0.16+0.01+0.32}_{-0.13-0.01-0.21}$&$1.5$&$-$&$-$\\
				$\mathcal{B} r(B_s\to D_{s1}^{'-}\pi^{+})$&$0.068^{+0.025+0.010+0.074}_{-0.020-0.007-0.049}$&$0.071^{+0.026+0.011+0.077}_{-0.021-0.007-0.051}$&$0.081^{+0.028+0.011+0.080}_{-0.022-0.007-0.055}$&$0.084^{+0.029+0.011+0.083}_{-0.023-0.007-0.057}$&$0.29$&$0.7$&$1.5$\\
				$\mathcal{B} r( B_s\to D_{s1}^{'-}\rho^{+})$&$0.20^{+0.07+0.02+0.18}_{-0.06-0.00-0.12}$&$0.21^{+0.08+0.06+0.19}_{-0.06-0.01-0.13}$&$0.23^{+0.08+0.06+0.21}_{-0.06-0.04-0.14}$&$0.24^{+0.08+0.02+0.21}_{-0.07-0.01-0.14}$&$0.83$&$1.9$&$3.8$\\
				$\mathcal{B} r(B_s\to D_{s1}^{'-}K^{+})$&$0.006^{+0.002+0.008+0.006}_{-0.002-0.005-0.004}$&$0.006^{+0.002+0.009+0.006}_{-0.002-0.006-0.004}$&$0.007^{+0.002+0.009+0.007}_{-0.002-0.006-0.004}$&$0.007^{+0.002+0.009+0.007}_{-0.002-0.006-0.005}$&$0.021$&$0.054$&$-$\\
				$\mathcal{B} r(B_s\to D_{s1}^{'-}K^{\ast+})$&$0.012^{+0.004+0.009+0.011}_{-0.004-0.004-0.007}$&$0.012^{+0.005+0.009+0.011}_{-0.004-0.004-0.007}$&$0.014^{+0.005+0.001+0.012}_{-0.004-0.005-0.008}$&$0.014^{+0.005+0.001+0.012}_{-0.004-0.005-0.008}$&$0.044$&$0.1$&$-$\\
				$\mathcal{B} r( B_s\to D_{s1}^{'-}D^{+})$&$0.009^{+0.003+0.001+0.010}_{-0.003-0.009-0.006}$&$0.009^{+0.003+0.001+0.009}_{-0.003-0.009-0.006}$&$0.011^{+0.004+0.002+0.011}_{-0.003-0.001-0.008}$&$0.011^{+0.004+0.001+0.011}_{-0.003-0.009-0.007}$&$-$&$-$&$-$\\
				$\mathcal{B} r( B_s\to D_{s1}^{'-}D^{\ast+})$&$0.018^{+0.007+0.001+0.013}_{-0.005-0.000-0.008}$&$0.019^{+0.007+0.001+0.013}_{-0.005-0.000-0.008}$&$0.021^{+0.007+0.001+0.015}_{-0.006-0.000-0.010}$&$0.023^{+0.007+0.001+0.015}_{-0.006-0.000-0.010}$&$-$&$-$&$-$\\
				\hline\hline
			\end{tabular}\label{mix state}}
	\end{center}
\end{table}
\item Our predictions for the branching ratios of the decays $B_s\to D_{s1} P(V)$ are at least one order larger than those of the corresponding decays $B_s\to D_{s1}^\prime P(V)$. This is because that the related form factor $V_0^{B_sD_{s1}}$ is much larger than that of $V_0^{B_sD_{s1}^\prime}$. There exists the similar situation between the branching ratios of the decays $B\to D^{(*)}D_{s1}$ and $B\to D^{(*)}D^\prime_{s1}$, where $D^{(\prime)}_{s1}$ is at the emission position in the Feynman diagrams.  
\item In view of the mixing angle $\theta_s$ uncertainty, we check the dependence of the branching ratios of the decays $B_s\to D_{s1}^{(\prime)}\pi(K)$ on the mixing angle $\theta_s$, which are shown in Figure \ref{fig:T3}. One can find that the branching ratios of the decays  $B_s\to D_{s1}^{\prime}\pi(K)$ are very sensitive to the mixing angle, while those of the decays $B_s\to D_{s1}\pi(K)$ show an insensitive dependence on $\theta_s$. Furthermore, the changing trends for the branching ratios of these two kinds of decays are just opposite.
 \end{itemize}
 	\begin{figure}[H]
	\vspace{0.32cm}
	\centering
	\subfigure[]{\includegraphics[width=0.38\textwidth]{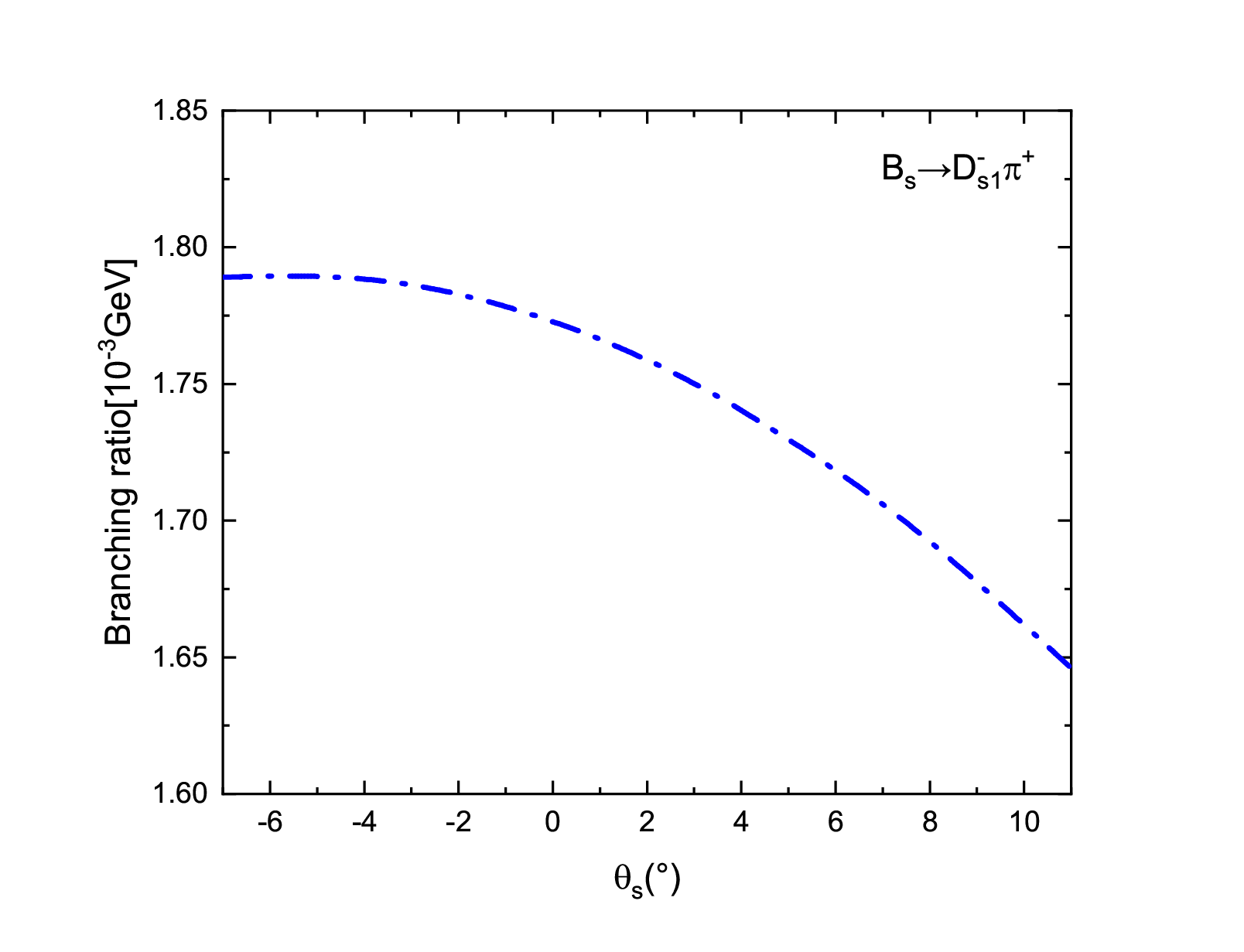}\quad}
	\subfigure[]{\includegraphics[width=0.38\textwidth]{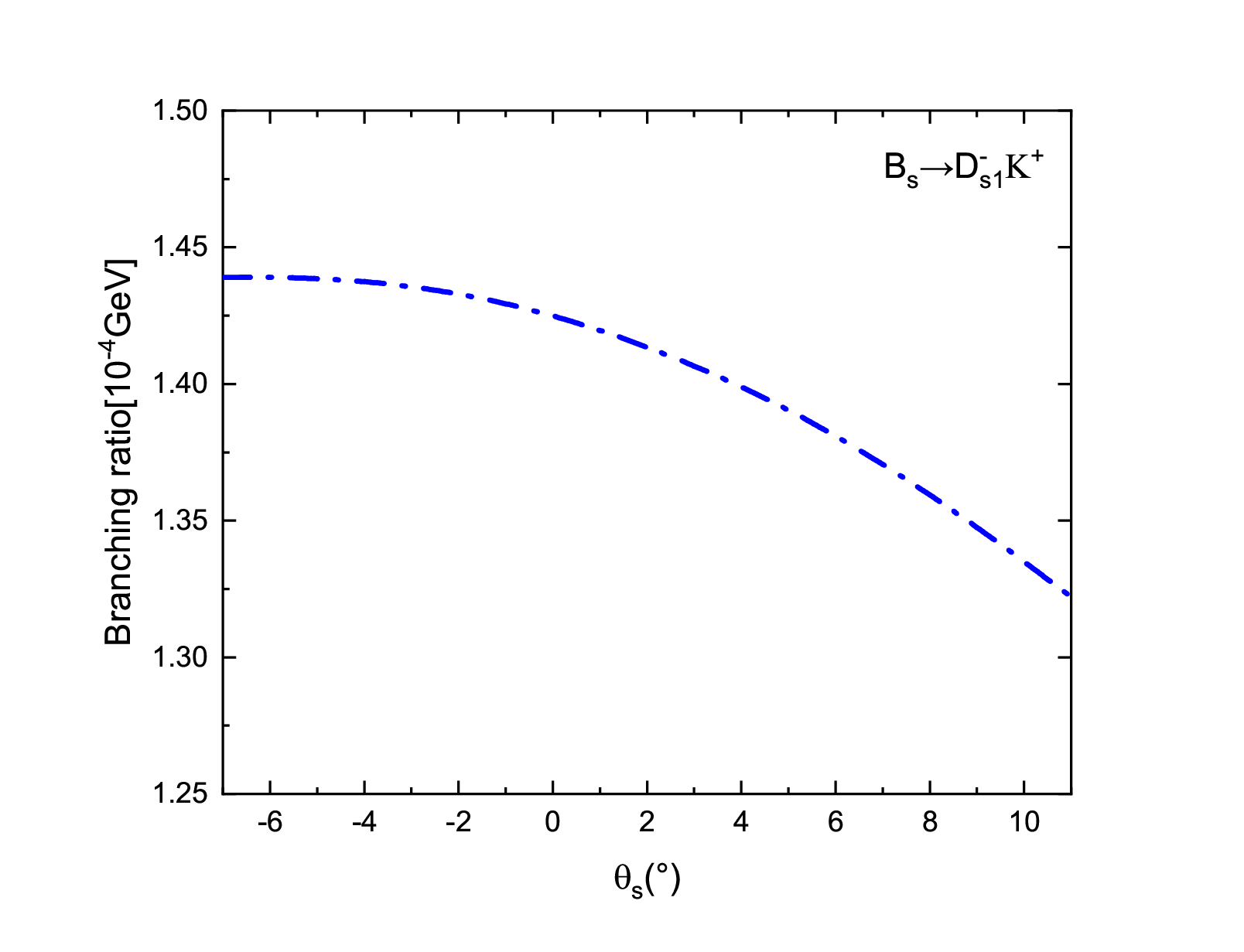}}\\
	\subfigure[]{\includegraphics[width=0.38\textwidth]{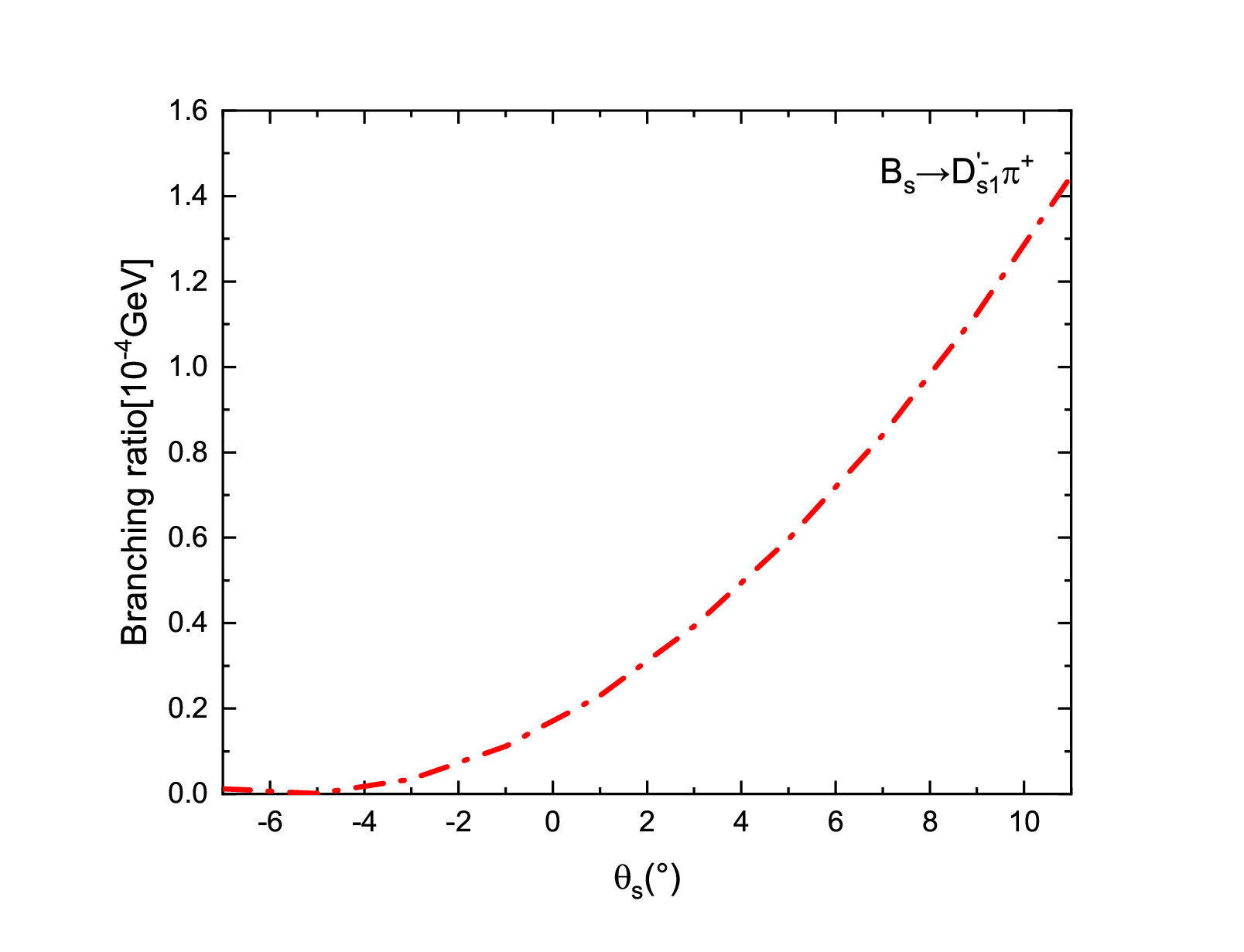}\quad}
	\subfigure[]{\includegraphics[width=0.38\textwidth]{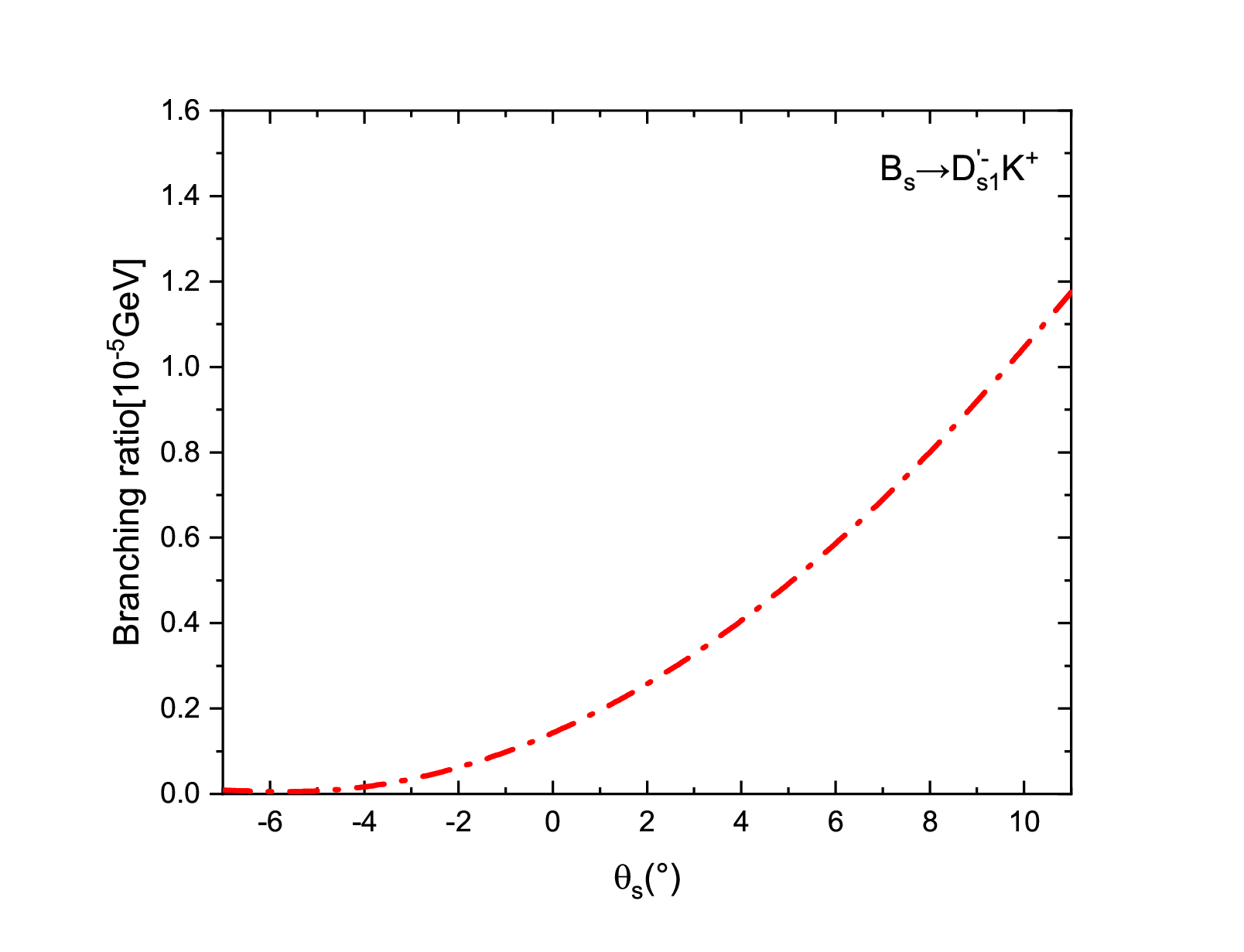}}\\
	\caption{The dependence of the branching ratios of the decays  $B_s\to D_{s1}^{(\prime)}\pi(K)$ on the mixing angle $\theta_s$. }\label{fig:T3}
\end{figure}

\begin{table}[H]
\caption{The branching ratios ($10^{-3}$) of the decays $B\to D^{(*)}D^{(\prime)}_{s1}$. The labels LO, VC, HSSC, NLO and the error sources are the same with those given in Table \ref{D2300pik}. Other theoretical predictions and data are also listed for comparison.  }
\begin{center}
\scalebox{0.6}{
\begin{tabular}{|c|c|c|c|c|c|c|c|c|c|}
\hline\hline
&LO& VC&HSSC&NLO &ISGW2 \cite{Cheng:2003id}&ISGW \cite{Katoch:1996hr} &FH \cite{Faessler:2007cu}&TM \cite{Liu:2022dmm}&Data \cite{pdg22, ijmp}\\
\hline\hline
$\mathcal{B} r( B^0\to D^{-} D_{s1}^+)$&$6.41^{+0.02+0.09}_{-0.10-0.12}$&$7.08^{+0.02+0.09}_{-0.11-0.13}$&$6.33^{+0.02+0.08}_{-0.10-0.12}$&$6.99^{+0.02+0.09}_{-0.11-0.13}$&$3.9$&$3.4$&$2.36\pm0.36$&$1.158\pm0.324$&$3.5\pm1.1$\\
$\mathcal{B} r( B^0\to D^{\ast-} D_{s1}^+)$&$9.78^{+0.08+0.52}_{-0.20-0.58}$&$10.52^{+0.09+0.57}_{-0.22-0.63}$&$9.70^{+0.08+0.52}_{-0.20-0.58}$&$10.44^{+0.08+0.56}_{-0.22-0.63}$&$15$&$-$&$6.85\pm1.05$&$2.709\pm0.759$&$9.3\pm2.2$\\
$\mathcal{B} r( B^+\to \bar{D}^{0}
D_{s1}^+)$&$6.92^{+0.02+0.09}_{-0.10-0.13}$&$7.64^{+0.02+0.10}_{-0.12-0.14}$&$6.83^{+0.02+0.09}_{-0.10-0.13}$&$7.54^{+0.02+0.10}_{-0.11-0.14}$&$4.3$&$3.5$&$2.54\pm0.39$&$1.255\pm0.351$&$3.1^{+1.0}_{-0.9}$\\
$\mathcal{B} r(B^+\to \bar{D}^{\ast0}D_{s1}^+)$&$10.55^{+0.09+0.56}_{-0.22-0.63}$&$11.35^{+0.09+0.61}_{-0.24-0.68}$&$10.46^{+0.08+0.56}_{-0.22-0.62}$&$11.26^{+0.09+0.61}_{-0.23-0.67}$&$16$&$-$&$7.33\pm1.12$&$3.065\pm0.858$&$12.0\pm3.0$\\
\hline
$\mathcal{B} r( B^0\to D^{-} D_{s1}^{'+})$&$0.29^{+0.00+0.00}_{-0.00-0.01}$&$0.32^{+0.00+0.00}_{-0.00-0.01}$&$0.29^{+0.00+0.00}_{-0.00-0.01}$&$0.32^{+0.00+0.00}_{-0.00-0.01}$&$0.28$&$3.3$&$-$&$-$&$0.39\pm0.18^a$\\
$\mathcal{B} r( B^0\to D^{\ast-} D_{s1}^{'+})$&$0.45^{+0.01+0.02}_{-0.00-0.03}$&$0.48^{+0.01+0.03}_{-0.00-0.03}$&$0.44^{+0.00+0.02}_{-0.01-0.03}$&$0.48^{+0.00+0.03}_{-0.01-0.03}$&$1.1$&$-$&$-$&$-$&$0.71\pm0.28^b$\\
$\mathcal{B} r(B^+\to \bar{D}^{0}D_{s1}^{'+})$&$0.32^{+0.00+0.00}_{-0.00-0.01}$&$0.35^{+0.01+0.00}_{-0.00-0.01}$&$0.31^{+0.00+0.00}_{-0.00-0.01}$&$0.34^{+0.00+0.00}_{-0.00-0.01}$&$0.31$&$3.4$&$-$&$-$&$0.35\pm0.16^a$\\
$\mathcal{B} r(B^+\to \bar{D}^{\ast0} D_{s1}^{'+})$&$0.48^{+0.00+0.03}_{-0.01-0.03}$&$0.52^{+0.00+0.03}_{-0.01-0.03}$&$0.48^{+0.00+0.03}_{-0.01-0.03}$&$0.51^{+0.00+0.03}_{-0.01-0.03}$&$1.2$&$-$&$-$&$-$&$0.91\pm0.36^b$\\
\hline\hline
\end{tabular}\label{last}}\\
\end{center}
{\tiny $^a$ It is obtained from the ratios $\frac{\mathcal{B}r(B\to DD_{s1})}{\mathcal{B}r(B\to DD^*_{s})}=0.44\pm0.11$ and $\frac{\mathcal{B}r(B\to DD^\prime_{s1})}{\mathcal{B}r(B\to DD^*_{s})}=0.049\pm0.010$ \cite{ijmp}.}\\
{\tiny $^b$ It is obtained from the ratios $\frac{\mathcal{B}r(B\to D^*D_{s1})}{\mathcal{B}r(B\to D^*D^*_{s})}=0.58\pm0.12$ and $\frac{\mathcal{B}r(B\to D^*D^\prime_{s1})}{\mathcal{B}r(B\to D^*D^*_{s})}=0.044\pm0.010$ \cite{ijmp}.}\\
\end{table}

In Table \ref{last}, we present our predictions for the branching ratios of the decays $B\to D^{(*)}D^{(\prime)}_{s1}$, which are associated with the $B\to D^{(*)}$ transition, accompanied by the $D^{(')}_{s1}$ emission. When the emission meson is $D_{s1}$, our results for the decays $B^0\to D^- D_{s1}^+$ and $B^+\to \bar{D}^0 D_{s1}^+$ are larger than those given by the ISGW(2) \cite{Cheng:2003id, Katoch:1996hr} and the FH \cite{Faessler:2007cu}
 by a factor of $2$ to $3$. While for other two decays $B^0\to D^{*-} D_{s1}^+$ and $B^+\to \bar{D}^{*0} D_{s1}^+$, our predictions are consistent well with the theoretical and experimental results within errors except for those given in Ref. \cite{Liu:2022dmm}, where the triangle mechanism was used by considering $D_{s1}$ as a molecular state.
When the emission meson is $D_{s1}^{\prime}$, the branching ratios of the decays $B^0\to D^{(*)-} D_{s1}^{\prime+}$ and $B^+\to \bar{D}^{(*)0} D_{s1}^{\prime+}$ are at least one order smaller than those of the decays $B^0\to D^{(*)-} D_{s1}^+$ and $B^+\to \bar{D}^{(*)0} D_{s1}^+$. It is because of the much smaller decay constant $f_{D^\prime_{s1}}$ compared to $f_{D_{s1}}$. Such character has been verified by the data shown in Table \ref{last}. 
\subsection{SUMMARY}
%%%%%%%%%%%%%%%%%%%%%%%%%%%%%%%%%%%%%%%%%%%%%%%%%%%%%%%%%%%%%%%%%%%%%%%%%%%%%%%
Firstly, we studied the form factors of the transitions  $B_{(s)}\to D_{0}^{\ast}, D_{s0}^{\ast},$ $D_{s1}$ and $D_{s1}^\prime$ in the covariant light-front quark model (CLFQM). One can find that these form factors are (much) smaller than those of the transitions $B_{(s)}\to D_{(s)}, D^*_{(s)}$. Certainly, because of the mixing between $D_{s1}$ and $D_{s1}^\prime$, the determination of the form factors for the transitions $B_{s}\to D_{s1}, D_{s1}^\prime$ are more difficult. Secondly, using the amplitudes combined via the form factors, we calculated the branching ratios of the $B_{(s)}$ nonleptonic decays with these four charmed mesons involved. Furthermore, the QCD radiative corrections to the hadronic matrix elements within the framework of QCD factorization were included. From the numerical results, we found the following points
\begin{enumerate}
\item The small form factors of the transitions $B_{(s)}\to D_{0}^{\ast}, D_{s0}^{\ast}$ are related to the small decay constants $f_{D^*_0}, f_{D^*_{s0}}$. Unfortunately, there exit large uncertainties in these two decay constants.
Combined with the data,	our predictions for the branching ratios of the $B_{(s)}$ decays with $D_{0}^{\ast}, D_{s0}^{\ast}$ involved are helpful to probe the inner structures of these two states. 
Most of the decays $B_{(s)}\to D_{0}^{\ast}P(V), D_{s0}^{\ast}P(V)$ with P(V) being a light pseudoscalar (vector) meson or a charmed meson are not sensitive to the QCD radiative corrections including the vertex corrections and the hard spectator-scattering, except for the decays $B^+\to \bar{D}_{0}^{\ast0}\pi^+(K^+)$, where two kinds of Feynman diagrams contribute to the branching ratios.
Our predictions for the branching ratios of the quasi-two-body decays $B^{+}\to \bar{D}^{\ast0}_{0}\pi^{+}\to D^{-}\pi^{+}\pi^{+}$ and ${B}^{0}\to D^{\ast-}_{0}K^{+}\to \bar{D}^{0}\pi^{-}K^{+}$ can explain the data by taking appropriate decay constant value for $D^{\ast}_{0}$, while for the decays $B^{+}\to \bar{D}^{\ast0}_{0} K^{+}\to D^{-}\pi^{+}K^{+}$ and ${B}^{0}\to D^{\ast-}_{0}\pi^{+}\to \bar{D}^{0}\pi^{-}\pi^{+}$, their branching ratios are much larger than the LHCb measurements. There exist the similar cases for the PQCD calculations compared with the data.  

\item We checked the dependence of the branching ratios of the decays $B_s\to D_{s1}P(V), D_{s1}^\prime P(V)$ on the mixing angle $\theta_s$ and found that the branching ratios of the decays $B_s\to D_{s1}^\prime P(V)$ are very sensitive to the mixing angle, while those of the decays $B_s\to D_{s1} P(V)$ show an insensitive dependence on $\theta_s$. The changing trends of the branching ratios between these two kinds of decays are just opposite. Furthermore, the branching ratios of the decays $B_s\to D_{s1}P(V)$ are at least one order larger than those of the decays $B_s\to D_{s1}^\prime P(V)$. It is because of the larger form factor $V^{B_sD_{s1}}_0$ compared to $V^{B_sD_{s1}^\prime}_0$. Such a feature is similar to the decays $B\to D^{(*)}D^{(\prime)}_{s1}$, where the $D^{(\prime)}_{s1}$ is at the emission position in the Feynman diagrams, that is to say the branching ratios of the decays $B\to D^{(*)} D_{s1}$ are at least one order larger than those of the decays $B\to D^{(*)} D_{s1}^{\prime}$. It is because of the larger decay constant $f_{D_{s1}}$ compared to $f_{D_{s1}^\prime}$. This character has been verified by the experimental measurements. 

\item Our predictions are helpful to clarify the different assumptions about the inner structures of these four charmed hadrons by comparing with the future data.
\end{enumerate}
\section*{Acknowledgment}
This work is partly supported by the National Natural Science
Foundation of China under Grant No. 11347030, by the Program of
Science and Technology Innovation Talents in Universities of Henan
Province 14HASTIT037.
\appendix
\section{Some specific rules under the $p^-$ intergration}
When preforming the integraion, we need to include the zero-mode contributions. It amounts to performing the integration in a proper way in the CLFQM. Specificlly we
use the following rules given in Refs. \cite{Jaus:1999zv,Cheng:2003sm}
\be \hat{p}_{1 \mu}^{\prime} &\doteq &   P_{\mu}
A_{1}^{(1)}+q_{\mu} A_{2}^{(1)},\\
\hat{p}_{1 \mu}^{\prime}
\hat{p}_{1 \nu}^{\prime}  &\doteq & g_{\mu \nu} A_{1}^{(2)} +P_{\mu}
P_{\nu} A_{2}^{(2)}+\left(P_{\mu} q_{\nu}+q_{\mu} P_{\nu}\right)
A_{3}^{(2)}+q_{\mu} q_{\nu} A_{4}^{(2)},\\
Z_{2}&=&\hat{N}_{1}^{\prime}+m_{1}^{\prime 2}-m_{2}^{2}+\left(1-2
x_{1}\right) M^{\prime 2} +\left(q^{2}+q \cdot P\right)
\frac{p_{\perp}^{\prime} \cdot q_{\perp}}{q^{2}},
\en
\be
A_{1}^{(1)}&=&\frac{x_{1}}{2}, \quad A_{2}^{(1)}=
A_{1}^{(1)}-\frac{p_{\perp}^{\prime} \cdot q_{\perp}}{q^{2}},\quad A_{3}^{(2)}=A_{1}^{(1)} A_{2}^{(1)},\\
A_{4}^{(2)}&=&\left(A_{2}^{(1)}\right)^{2}-\frac{1}{q^{2}}A_{1}^{(2)},\quad A_{1}^{(2)}=-p_{\perp}^{\prime 2}-\frac{\left(p_{\perp}^{\prime}
\cdot q_{\perp}\right)^{2}}{q^{2}}, \quad A_{2}^{(2)}=\left(A_{1}^{(1)}\right)^{2}.  \en

\section{EXPRESSIONS OF $B \rightarrow D^*_0, \;^iD_{s1}$ FORM FACTORS}
\begin{footnotesize}
\begin{eqnarray}
F_{1}^{B D^*_0}\left(q^{2}\right)  &=& \frac{N_{c}}{16 \pi^{3}} \int d x_{2} d^{2} p_{\perp}^{\prime} \frac{h_{B}^{\prime}
	h_{D^*_0}^{\prime \prime}}{x_{2} \hat{N}_{1}^{\prime} \hat{N}_{1}^{\prime \prime}}\left[x_{1}\left(M_{0}^{\prime 2}+M_{0}^{\prime \prime 2}\right)+x_{2} q^{2}\right.\non &&
\left.-x_{2}\left(m_{1}^{\prime}+m_{1}^{\prime
	\prime}\right)^{2}-x_{1}\left(m_{1}^{\prime}-m_{2}\right)^{2}-x_{1}\left(m_{1}^{\prime
	\prime}+m_{2}\right)^{2}\right],\\
F_{0}^{B D^*_0}\left(q^{2}\right) &=& F_{1}^{B D^*_0}\left(q^{2}\right)+\frac{q^{2}}{q \cdot P} \frac{N_{c}}{16 \pi^{3}} \int d x_{2} d^{2} p_{\perp}^{\prime}
\frac{2 h_{B}^{\prime} h_{D^*_0}^{\prime \prime}}{x_{2} \hat{N}_{1}^{\prime} \hat{N}_{1}^{\prime \prime}}\left\{-x_{1} x_{2} M^{\prime 2}
-p_{\perp}^{\prime 2}-m_{1}^{\prime} m_{2}\right.\non &&\left.
-\left(m_{1}^{\prime \prime}+m_{2}\right)\left(x_{2} m_{1}^{\prime}+x_{1} m_{2}\right)
+2 \frac{q \cdot P}{q^{2}}\left(p_{\perp}^{\prime 2}+2 \frac{\left(p_{\perp}^{\prime} \cdot q_{\perp}\right)^{2}}{q^{2}}\right)
+2 \frac{\left(p_{\perp}^{\prime} \cdot q_{\perp}\right)^{2}}{q^{2}}\right.\non &&\left.
-\frac{p_{\perp}^{\prime} \cdot q_{\perp}}{q^{2}}\left[M^{\prime \prime 2}-x_{2}\left(q^{2}+q \cdot P\right)
-\left(x_{2}-x_{1}\right) M^{\prime 2}+2 x_{1} M_{0}^{\prime 2}\right.\right.\non &&\left.\left.-2\left(m_{1}^{\prime}-m_{2}\right)
\left(m_{1}^{\prime}-m_{1}^{\prime \prime}\right)\right]\right\},\\
A^{B\;^iD_{s1}}(q^{2})&=&(M^{'}-M^{''})\frac{N_{c}}{16 \pi^{3}} \int d x_{2} d^{2} p_{\perp}^{\prime} \frac{2 h_{B}^{\prime}
 h_{\;^iD_{s1}}^{\prime \prime}}{x_{2} \hat{N}_{1}^{\prime} \hat{N}_{1}^{\prime \prime}}\left\{x_{2} m_{1}^{\prime}
 +x_{1} m_{2}+\left(m_{1}^{\prime}+m_{1}^{\prime \prime}\right) \frac{p_{\perp}^{\prime} \cdot q_{\perp}}{q^{2}}\right.\non &&\left.
 +\frac{2}{w_{^iD_{s1}}^{\prime \prime}}\left[p_{\perp}^{\prime 2}+\frac{\left(p_{\perp}^{\prime} \cdot q_{\perp}\right)^{2}}{q^{2}}\right]\right\},\\
V_1^{B\;^iD_{s1}}(q^{2})&=& -\frac{1}{M^{'}-M^{''}}\frac{N_{c}}{16 \pi^{3}} \int d x_{2} d^{2} p_{\perp}^{\prime} \frac{h_{B}^{\prime} h_{\;^iD_{s1}}^{\prime \prime}}{x_{2}
\hat{N}_{1}^{\prime}
\hat{N}_{1}^{\prime \prime}}\{2 x_{1}\left(m_{2}-m_{1}^{\prime}\right)\left(M_{0}^{\prime 2}+M_{0}^{\prime \prime 2}\right)
+4 x_{1} m_{1}^{\prime \prime} M_{0}^{\prime 2}\non&&+2 x_{2} m_{1}^{\prime} q \cdot P
\left.+2 m_{2} q^{2}-2 x_{1} m_{2}\left(M^{\prime 2}+M^{\prime \prime 2}\right)+2\left(m_{1}^{\prime}-m_{2}\right)\left(m_{1}^{\prime}
-m_{1}^{\prime \prime}\right)^{2}+8\left(m_{1}^{\prime}-m_{2}\right) \right.\non &&
\left. \times\left[p_{\perp}^{\prime 2}+\frac{\left(p_{\perp}^{\prime}
\cdot q_{\perp}\right)^{2}}{q^{2}}\right]+2\left(m_{1}^{\prime}-m_{1}^{\prime \prime}\right)\left(q^{2}+q \cdot P\right) \frac{p_{\perp}^{\prime} \cdot q_{\perp}}{q^{2}}
-4 \frac{q^{2} p_{\perp}^{\prime 2}+\left(p_{\perp}^{\prime} \cdot q_{\perp}\right)^{2}}{q^{2} w_{^iD_{s1}}^{\prime \prime}}
\right.\non && \left.\times\left[2 x_{1}\left(M^{\prime 2}+M_{0}^{\prime 2}\right)-q^{2}-q \cdot P-2\left(q^{2}+q \cdot P\right) \frac{p_{\perp}^{\prime} \cdot q_{\perp}}{q^{2}}-2\left(m_{1}^{\prime}+m_{1}^{\prime \prime}\right)\left(m_{1}^{\prime}-m_{2}\right)\right]\right\},\;\;\;\;\;\;\;\\
V_2^{B \;^iD_{s1}}(q^{2})&=& (M^{'}-M^{''})\frac{N_{c}}{16 \pi^{3}} \int d x_{2} d^{2} p_{\perp}^{\prime} \frac{2 h_{B}^{\prime} h_{\;^iD_{s1}}^{\prime \prime}}{x_{2} \hat{N}_{1}^{\prime}
\hat{N}_{1}^{\prime \prime}}\left\{(x_{1}-x_{2}\right)\left(x_{2} m_{1}^{\prime}+x_{1} m_{2}\right)-[2 x_{1} m_{2}
-m_{1}^{\prime \prime}\non &&+\left(x_{2}-x_{1}\right) m_{1}^{\prime}]
\times \frac{p_{\perp}^{\prime} \cdot q_{\perp}}{q^{2}}-2 \frac{x_{2} q^{2}+p_{\perp}^{\prime} \cdot q_{\perp}}{x_{2} q^{2} w_{^iD_{s1}}^{\prime \prime}}[p_{\perp}^{\prime} \cdot p_{\perp}^{\prime \prime}
+\left(x_{1} m_{2}+x_{2} m_{1}^{\prime}\right)\non &&\times\left(x_{1} m_{2}+x_{2} m_{1}^{\prime \prime}\right)]\},
\end{eqnarray}
\end{footnotesize}
\begin{footnotesize}
\begin{eqnarray}
V_0^{B \;^iD_{s1}}(q^{2})&=& \frac{M^{'}-M^{''}}{2M^{''}}V_1^{B \;^iD_{s1}}(q^{2})-\frac{M^{'}+M^{''}}{2M^{''}}V_2^{B \;^iD_{s1}}(q^{2})-\frac{q^2}{2M^{''}}\frac{N_{c}}{16 \pi^{3}} \int d x_{2} d^{2} p_{\perp}^{\prime} \frac{h_{B}^{\prime} h_{\;^iD_{s1}}^{\prime \prime}}{x_{2} \hat{N}_{1}^{\prime}
\hat{N}_{1}^{\prime \prime}}\non &&\times\{2\left(2 x_{1}-3\right)\left(x_{2} m_{1}^{\prime}+x_{1} m_{2}\right)-8\left(m_{1}^{\prime}-m_{2}\right)
\left[\frac{p_{\perp}^{\prime 2}}{q^{2}}
+2 \frac{\left(p_{\perp}^{\prime} \cdot q_{\perp}\right)^{2}}{q^{4}}\right]-[\left(14-12 x_{1}\right) m_{1}^{\prime}\non &&+2 m_{1}^{\prime \prime}-\left(8-12 x_{1}\right) m_{2}] \frac{p_{\perp}^{\prime} \cdot q_{\perp}}{q^{2}}
+\frac{4}{w_{^iD_{s1}}^{\prime \prime}}(\left[M^{\prime 2}+M^{\prime \prime 2}-q^{2}+2\left(m_{1}^{\prime}-m_{2}\right)\left(-m_{1}^{\prime \prime}
+m_{2}\right)\right]\non &&\times\left(A_{3}^{(2)}+A_{4}^{(2)}-A_{2}^{(1)}\right)
+Z_{2}\left(3 A_{2}^{(1)}-2 A_{4}^{(2)}-1\right)+\frac{1}{2}[x_{1}\left(q^{2}+q \cdot P\right)
-2 M^{\prime 2}-2 p_{\perp}^{\prime} \cdot q_{\perp}\non &&-2 m_{1}^{\prime}\left(-m_{1}^{\prime \prime}+m_{2}\right)
\left.-2 m_{2}\left(m_{1}^{\prime}-m_{2}\right)\right]\left(A_{1}^{(1)}+A_{2}^{(1)}-1\right)\non &&
\left.\left.\times q \cdot P\left[\frac{p_{\perp}^{\prime 2}}{q^{2}}
+\frac{\left(p_{\perp}^{\prime} \cdot q_{\perp}\right)^{2}}{q^{4}}\right]\left(4 A_{2}^{(1)}-3\right)\right)\right\},\;\;\;
\end{eqnarray}
\end{footnotesize}
with $i=1,3$.

%%%%%%%%%%%%%%%%%%%%%%%%%%%%%%%%%%%%%%%%%%%%%%%%%%%%%%%%%%%%%%%%%%%%%%%%
%                               references
%%%%%%%%%%%%%%%%%%%%%%%%%%%%%%%%%%%%%%%%%%%%%%%%%%%%%%%%%%%%%%%%%%%%%%%%


\begin{thebibliography}{99}
\bibitem{babar}
B. Aubert \emph{et al}. [BaBar],
Phys. Rev. Lett. {\bf90}, 242001 (2003) [arXiv:hep-ex/0304021].
\bibitem{cleo}
D. Besson \emph{et al}. [CLEO], Phys. Rev. D {\bf68}, 032002 (2003) [Erratum: Phys. Rev. D {\bf75}, 119908 (2007)] [arXiv:hep-ex/0305100].
\bibitem{belle}
K. Abe, \emph{et al}. [Belle], Phys. Rev. D {\bf69} 112002 (2004) [arXiv:hep-ex/0307021].
\bibitem{nun}
A. E. Asratian, \emph{et al}., Z. Phys. C {\bf40} 483 (1988). 
\bibitem{quark}
S. Godfrey and N. Isgur, Phys. Rev. D {\bf32}, 189 (1985).
\bibitem{guo}
Z. X. Xie, G. Q. Feng and X. H. Guo, Phys. Rev. D {\bf81}, 036014 (2010).
\bibitem{cleven}
M. Cleven, H. W. Griehammer, F. K. Guo, C. Hanhart and U. G. Meiner, Eur. Phys. J. A {\bf50}, 149 (2014) [arXiv:1405.2242 [hep-ph]].
\bibitem{guofk}
F. K. Guo, P. N. Shen, H. C. Chiang, R. G. Ping, and B. S. Zou, Phys. Lett. B {\bf641}, 278 (2006) [arXiv:hep-ph/0603072].
\bibitem{close}
T. Barnes, F. E. Close and  H. J. Lipkin, Phys. Rev. D {\bf68}, 054006 (2003) [arXiv:hep-ph/0305025].
\bibitem{lutz}
 E. E. Kolomeitsev and M. F. M. Lutz, Phys. Lett. B {\bf582}, 39 (2004) [arXiv:hep-ph/0307133].
\bibitem{lutz2}
J. Hofmann and M. F. M. Lutz, Nucl. Phys. A {\bf733}, 142 (2004) [arXiv:hep-ph/0308263].
\bibitem{ylma}
C. J. Xiao, D. Y. Chen and Y. L. Ma, Phys. Rev. D {\bf93}, 094011 (2016) [arXiv:1601.06399 [hep-ph]].
\bibitem{maiani}
L. Maiani, F. Piccinini, A. D. Polosa, and V. Riquer, Phys. Rev. D {\bf71}, 014028 (2005) [arXiv:hep-ph/0412098].
\bibitem{wangzg}
Z. G. Wang and S. L. Wan, Nucl. Phys. A {\bf778}, 22 (2006) [arXiv:hep-ph/0602080].
\bibitem{hycheng2}
H. Y. Cheng and W. S. Hou, Phys. Lett. B {\bf566}, 193 (2003) [arXiv:hep-ph/0305038].
\bibitem{yqchen}
Y. Q. Chen and X. Q. Li, Phys. Rev. Lett. {\bf93}, 232001 (2004) [arXiv:hep-ph/0407062].
\bibitem{kim}
H. Kim and Y. Oh, Phys. Rev. D {\bf72}, 074012 (2005) [arXiv:hep-ph/0508251].
\bibitem{bardeen}
W. A. Bardeen, E. J. Eichten, and C. T. Hill, Phys. Rev. D {\bf68}, 054024 (2003) [arXiv:hep-ph/0305049].
\bibitem{nowak}
M. A. Nowak, M. Rho, and I. Zahed, Acta Phys. Polon. B {\bf35}, 2377 (2004) [arXiv:hep-ph/0307102].
\bibitem{browder}
T. E. Browder, S. Pakvasa, and A. A. Petrov, Phys. Lett. B {\bf578}, 365 (2004) [arXiv:hep-ph/0307054].
\bibitem{vijande}
J. Vijande, F. Fernandez, and A. Valcarce, Phys. Rev. D {\bf73}, 034002 (2006) [arXiv:hep-ph/0601143].
\bibitem{bracco}
M. E. Bracco, A. Lozea, R. D. Matheus, F. S. Navarra and M. Nielsen, Phys. Lett. B {\bf624}, 217 (2005) [arXiv:hep-ph/0503137].
\bibitem{lutz3}
M. F. M. Lutz and M. Soyeur, Prog. Part. Nucl. Phys. {\bf61}, 155 (2008).
\bibitem{Hwang}
D. S. Hwang and D. W. Kim, Phys. Lett. B {\bf601}, 137 (2004) [arXiv:hep-ph/0408154].
\bibitem{xliu}
X. Liu, Y. M. Yu, S. M. Zhao and X. Q. Li, Eur. Phys. J. C {\bf47}, 445 (2006) [arXiv:hep-ph/0601017].
\bibitem{xlchen}
J. Lu, X. L. Chen, W. Z. Deng and S. L. Zhu, Phys. Rev. D {\bf73}, 054012 (2006) [arXiv:hep-ph/0602167].
\bibitem{jbliu}
J. B. Liu and M. Z. Yang, JHEP {\bf1407}, 106 (2014) [arXiv:1307.4636 [hep-ph]].
\bibitem{wangzg2}
Z. G. Wang, Phys. Rev. D {\bf75}, 034013 (2007) [arXiv:hep-ph/0612225].
\bibitem{fajfer}
 S. Fajfer and A. P. Brdnik, Phys. Rev. D {\bf92} , 074047 (2015) [arXiv:1506.02716 [hep-ph]].
\bibitem{song}
Q. T. Song, D. Y. Chen, X. Liu and T. Matsuki, Phys. Rev. D {\bf91}, 054031 (2015) [arXiv:1501.03575 [hep-ph]].
\bibitem{sfchen}
S. F. Chen, J. Liu, H. Q. Zhou, and D. Y. Chen, Eur. Phys. J. C {\bf80}, 290 (2020) [arXiv:2003.07988 [hep-ph]].
\bibitem{babar2}
B. Aubert, \emph{et al}. [BaBar] , Phys. Rev. D {\bf79} 112004 (2009) [arXiv:0901.1291 [hep-ex]].
\bibitem{godfrey2}
S. Godfrey and R. Kokoski, Phys. Rev. D {\bf43} 1679 (1991).
\bibitem{nielsen}
M. Nielsen, R. D. Matheus, F. S. Navarra and M. E. Bracco, Nucl. Phys. B {\bf161} 193 (2006).
\bibitem{FOCUS:2003gru}
J. M. Link \textit{et al.} [FOCUS], Phys. Lett. B \textbf{586}, 11 (2004) [arXiv:hep-ex/0312060].
\bibitem{albala}
M. Albaladejo, P. Ferni\"{a}andez-Soler, F. K. Guo, and J. Nieves, Phys. Lett. B { \bf767}, 465 (2017) [arXiv:1610.06727 [hep-ph]].
\bibitem{mldu}
M. L. Du, M. Albaladejo, P. Ferni\"{a}andez-Soler, F. K. Guo, C. Hanhart, U. G. Meiner, J. Nieves, and D. L. Yao, Phys. Rev. D {\bf98}, 094018 (2018) [arXiv:1712.07957 [hep-ph]].
\bibitem{guofk2}
F. K. Guo, C. Hanhart, U. G. Meiner, Q. Wang, Q. Zhao, and B. S. Zou, Rev. Mod. Phys. {\bf90}, 015004 (2018) [arXiv:1705.00141 [hep-ph]].
\bibitem{mldu2}
M. L. Du, F. K. Guo, C. Hanhart, B. Kubis and U. G. Meiner, Phys. Rev. Lett. {\bf126}, 192001 (2021) [arXiv:2012.04599 [hep-ph]].
\bibitem{gamer}
D. Gamermann, E. Oset, D. Strottman and M. J. Vicente Vacas, Phys. Rev. D {\bf76}, 074016 (2007) [arXiv:hep-ph/0612179]
\bibitem{babar3}
B. Aubert, \emph{et al}. [BaBar], Phys. Rev. D {\bf77}, 011102 (2008) [arXiv:0708.1565 [hep-ex]].
\bibitem{belle2}
T. Aushev, \emph{et al}. [Belle], Phys. Rev. D {\bf83}, 051102 (2011) [arXiv:1102.0935 [hep-ex]].
\bibitem{lhcb}
R. Aaij, \emph{et al}. [LHCb], Phys. Rev. D {\bf86}, 112005 (2012) [arXiv:1211.1541 [hep-ex]].
\bibitem{belle3}
V. Balagura,\emph{et al}. [Belle], Phys. Rev. D {\bf77}, 032001 (2008) [arXiv:0709.4184 [hep-ex]].
\bibitem{zhw}
Z. H. Wang, Y. Zhang, T. h. Wang, Y. Jiang, Q. Li and G. L. Wang, Chin. Phys. C {\bf42}, 123101 (2018) [arXiv:1803.06822 [hep-ph]].
\bibitem{Salpeter:1951sz}
E. E. Salpeter and H. A. Bethe, Phys. Rev. {\bf84}, 1232 (1951).
\bibitem{Salpeter:1952ib}
E. E. Salpeter, Phys. Rev. {\bf87}, 328  (1952).
\bibitem{Jaus:1999zv}
W. Jaus, Phys. Rev. D {\bf60}, 054026 (1999).
\bibitem{Choi:1998nf}
H. M. Choi and C. R. Ji, Phys. Rev. D {\bf58}, 071901 (1998) [arXiv:hep-ph/9805438].
\bibitem{Cheng:1997au}
H. Y. Cheng, C. Y. Cheung, C. W. Hwang and W. M. Zhang, Phys. Rev. D {\bf57} 5598 (1998) [arXiv:hep-ph/9709412].
\bibitem{Cheng:2003sm}
H. Y. Cheng, C. K. Chua and C. W. Hwang, Phys. Rev. D {\bf69}, 074025 (2004) [arXiv:hep-ph/0310359].
\bibitem{w.wang}
 W. Wang, Y. L. Shen and C. D. Lu, Phys. Rev. D {\bf79}, 054012 (2009) [arXiv:0811.3748 [hep-ph]].
\bibitem{x.wang}
 X. X. Wang, W. Wang and C. D. Lu,  Phys. Rev. D {\bf79}, 114018 (2009) [arXiv:0901.1934 [hep-ph]].
\bibitem{w.wang2}
W. Wang, Y. L. Shen and C. D. Lu,  Eur. Phys. J. C {\bf51}, 841 (2007) [arXiv:0704.2493 [hep-ph]].
\bibitem{ke}
H. W. Ke, T. Liu and X. Q. Li,  Phys. Rev. D {\bf89}, 017501 (2014) [arXiv:1307.5925 [hep-ph]].
\bibitem{Li:2010bb}
G. Li, F. l. Shao and W. Wang, Phys. Rev. D \textbf{82}, 094031 (2010) [arXiv:1008.3696 [hep-ph]].
\bibitem{Sun1}
Z. Q. Zhang, Z. J. Sun, Y. C. Zhao, Y. Y. Yang and Z. Y. Zhang,
Eur. Phys. J. C \textbf{83}, 477 (2023) [arXiv:2301.11107 [hep-ph]].
\bibitem{Sun2}
Z. J. Sun, S. Y. Wang, Z. Q. Zhang, Y. Y. Yang and Z. Y. Zhang,
Eur. Phys. J. C \textbf{83}, 945 (2023) [arXiv:2308.03114 [hep-ph]].
\bibitem{Sun3}
Z. J. Sun, Z. Q. Zhang, Y. Y. Yang and H. Yang,
Eur. Phys. J. C \textbf{84}, 65 (2024) [arXiv:2311.04431 [hep-ph]].
\bibitem{bsw}
M. Wirbel, B. Stech and M. Bauer, Z. Phys. C {\bf29}, 637 (1985).
\bibitem{BBNS}
M. Beneke, G. Buchalla, M. Neubert, and C. T. Sachrajda, Phys. Rev. Lett. {\bf83}, 1914 (1999) [arXiv:hep-ph/9905312].
\bibitem{Cheng:2007st}
H. Y. Cheng, C. K. Chua and K. C. Yang,
Phys. Rev. D \textbf{77}, 014034 (2008) [arXiv:0705.3079 [hep-ph]].
\bibitem{BN}
 M. Beneke and M. Neubert, Nucl. Phys. B {\bf675}, 333 (2003) [arXiv:hep-ph/0308039].
\bibitem{pdg22}
R. L. Workman \textit{et al.} [Particle Data Group], Review of Particle Physics. PTEP {\bf2022}, 083C01 (2022).
\bibitem{Becirevic:1998ua}
D. Becirevic, P. Boucaud, J. P. Leroy, V. Lubicz, G. Martinelli, F. Mescia and F. Rapuano, Phys. Rev. D \textbf{60}, 074501 (1999) [arXiv:hep-lat/9811003].
\bibitem{Cheng:2003id}
H. Y. Cheng, Phys. Rev. D \textbf{68}, 094005 (2003) [arXiv:hep-ph/0307168].
\bibitem{Li:2009wq}
R. H. Li and C. D. Lu, Phys. Rev. D \textbf{80}, 014005 (2009) [arXiv:0905.3259 [hep-ph]].
\bibitem{T.M.}
T. M. Aliev and M. Savci, Phys. Rev. D {\bf73}, 114010 (2006) [arXiv:hep-ph/0604002].
\bibitem{Melikhov:2000yu}
D. Melikhov and B. Stech, Phys. Rev. D \textbf{62}, 014006 (2000) [arXiv:hep-ph/0001113].
\bibitem{Kramer2}
R. N. Faustov and V. O. Galkin, Phys. Rev. D \textbf{87}, 034033 (2013) [arXiv:1212.3167 [hep-ph]].
\bibitem{Kramer:1992xr}
G. Kramer and W. F. Palmer, Phys. Rev. D \textbf{46}, 3197 (1992).
\bibitem{Chen:2011ut}
X. J. Chen, H. F. Fu, C. S. Kim and G. L. Wang, J. Phys. G \textbf{39}, 045002 (2012) [arXiv:1106.3003 [hep-ph]].
\bibitem{Blasi:1993fi}
P. Blasi, P. Colangelo, G. Nardulli and N. Paver,
Phys. Rev. D \textbf{49}, 238 (1994) [arXiv:hep-ph/9307290].
\bibitem{Ball:1998tj}
P. Ball, JHEP \textbf{09}, 005 (1998) [arXiv:hep-ph/9802394].
\bibitem{Beneke:2001ev}
M. Beneke, G. Buchalla, M. Neubert and C. T. Sachrajda, Nucl. Phys. B \textbf{606}, 245 (2001) [arXiv:hep-ph/0104110].
\bibitem{Wang:2018fai}
W. F. Wang, Phys. Lett. B \textbf{788}, 468 (2019) [arXiv:1809.02943 [hep-ph]].
\bibitem{Katoch:1995hr}
A. C. Katoch and R. C. Verma, Int. J. Mod. Phys. A {\bf11}, 129 (1996).
\bibitem{LHCb:2016lxy}
R. Aaij \textit{et al.} [LHCb], Phys. Rev. D \textbf{94}, 072001 (2016) [arXiv:1608.01289 [hep-ex]].
\bibitem{Belle:2006wbx}
A. Kuzmin \textit{et al.} [Belle],
Phys. Rev. D \textbf{76}, 012006 (2007) [arXiv:hep-ex/0611054].
\bibitem{LHCb:2015klp}
R. Aaij \textit{et al.} [LHCb], Phys. Rev. D \textbf{92}, 032002 (2015) [arXiv:1505.01710 [hep-ex]].
\bibitem{LHCb:2015eqv}
R. Aaij \textit{et al.} [LHCb], Phys. Rev. D \textbf{91}, 092002 (2015) [erratum: Phys. Rev. D \textbf{93}, 119901 (2016)] [arXiv:1503.02995 [hep-ex]].
\bibitem{LHCb:2015tsv}
R. Aaij \textit{et al.} [LHCb],
Phys. Rev. D \textbf{92}, 012012 (2015) [arXiv:1505.01505 [hep-ex]].
\bibitem{Chua:2003ac}
C. K. Chua, J. Korean Phys. Soc. \textbf{45}, S256 (2004) [arXiv:hep-ph/0312075].
\bibitem{Zhang:2019pax}
Z. Q. Zhang, H. Guo, N. Wang and H. T. Jia, Phys. Rev. D \textbf{99}, 073002 (2019) [arXiv:1903.03990 [hep-ph]].
\bibitem{Albertus:2014bfa}
C. Albertus, Phys. Rev. D \textbf{89}, 065042 (2014) [arXiv:1401.1791 [hep-ph]].
\bibitem{Zhang:2021bcr}
Z. Q. Zhang, M. G. Wang, Y. C. Zhao, Z. L. Guan and N. Wang, Phys. Rev. D \textbf{103}, 116030 (2021) [arXiv:2105.02688 [hep-ph]].
\bibitem{Faessler:2007cu}
A. Faessler, T. Gutsche, S. Kovalenko and V. E. Lyubovitskij, Phys. Rev. D \textbf{76}, 014003 (2007) [arXiv:0705.0892 [hep-ph]].
\bibitem{Liu:2022dmm}
M. Z. Liu, X. Z. Ling, L. S. Geng, E. Wang and J. J. Xie, Phys. Rev. D \textbf{106}, 114011 (2022) [arXiv:2209.01103 [hep-ph]].
\bibitem{ijmp}
J. Segovia, D. R. Entem, F. Fernandez, E. Hernandez, Int. J. Mod. Phys. E {\bf22}, 1330026 (2013) [arXiv:1309.6926 [hep-ph]].
\bibitem{Katoch:1996hr}
A. C. Katoch and R. C. Verma, J. Phys. G {\bf22}, 1765 (1996).
\end{thebibliography}
\end{document}